\documentclass[aps,prb,twocolumn,showpacs,superscriptaddress,floatfix,longbibliography]{revtex4-2}
\usepackage{amssymb,amsmath}
\usepackage{graphicx}
\usepackage{multirow}
\usepackage{mathtools}
\usepackage{xcolor}
\usepackage{siunitx}
\usepackage{yfonts}
\usepackage{esint}
\usepackage{placeins}
\usepackage{tikz}
\usepackage[a4paper,margin=1in]{geometry}

\usepackage[colorlinks=true,allcolors=blue]{hyperref}

\graphicspath{{./IMG/}}

\usetikzlibrary{arrows.meta,positioning}

\begin{document}


\title{Capillarity in Stationary Random Granular Media: Distribution-Aware Screening and Quantitative Supercell Sizing}

\author{Christian Tantardini}
\email{christiantantardini@ymail.com}
\affiliation{Center for Integrative Petroleum Research, King Fahd University of Petroleum and Minerals, Dhahran 31261, Saudi Arabia}

\author{Fernando Alonso-Marroqu\'in}
\email{Alonsomarroquin@ibf.baugh.ethz.ch}
\affiliation{Center for Integrative Petroleum Research, King Fahd University of Petroleum and Minerals, Dhahran 31261, Saudi Arabia}
\affiliation{Department of Computational Physics for Engineering Material, ETH Zurich, 8092 Zurich, Switzerland}

\date{\today}

\begin{abstract}
We develop a quantitative framework to determine the minimal periodic supercell required for representative simulations of capillarity-screened Darcy flow in stationary random, polydisperse granular media. The microstructure is characterized by two-point statistics (covariance and spectral density) that govern finite-size fluctuations. Capillarity is modeled as a screened, modified-Helmholtz problem with phase-dependent transport under periodic boundary conditions; periodic homogenization yields an apparent conductivity, an apparent screening parameter, and a macroscopic capillary decay length. Because screening imparts a spatial low-pass response, we introduce a distribution-aware treatment of polydispersity consisting of a capillarity-weighted volume fraction and a screened analogue of the integral range that preserves variance units and recovers classical descriptors in the appropriate limits. These descriptors lead to two sizing rules: (i) a length criterion on the shortest cell edge controlled by a microstructural correlation length, the macroscopic decay length, and a high quantile of grain size; and (ii) a volume criterion that links the target coefficient of variation to the screened integral range and the phase contrast. The framework couples statistical microstructure information to capillary response and yields reproducible, distribution-aware supercell selection for image-based finite-element or fast-Fourier-transform solvers. The resulting criteria are therefore intended for representativity of the coarse-grained screened response, rather than for isolated nonlinear pore-scale events.
\end{abstract}

\maketitle

\section{Introduction}
\label{sec:introduction}

Capillarity in granular and porous materials controls drainage and imbibition, wetting front motion, transport, and even mechanical stability across many geological and engineered systems. When the grains are polydisperse and spatially disordered, the resulting pore–throat network spans multiple length scales and directions, so the capillary response is both \emph{screened} (it decays over a finite length) and \emph{statistical} (it depends on the sampled volume). In simulations, one typically replaces the infinite medium by a periodic supercell (hypertorus) and solves a periodic boundary-value problem. The practical question is simple but consequential: \emph{how large must the supercell be to obtain representative and reproducible capillary predictions?} Existing guidance is scattered—often problem-specific or based on ad-hoc multiples of grain size—rather than grounded in a quantitative link between microstructure statistics and capillary physics.

We address this gap using the statistical framework of random heterogeneous media \cite{Torquato2002,Chiu2013,DaleyVJ2003}. The microstructure is modeled as a stationary marked point process of grain centers and sizes; the associated two-phase indicator field yields standard two-point descriptors—the covariance and its spectral density—that control finite-size fluctuations. A single scalar summary, the \emph{integral range} (a correlation volume), emerges as the key parameter governing variance with supercell volume; its definition and variance implications are reviewed in \S\ref{subsec:integral_range_variance}, building on \cite{LuTorquato1990,QuintanillaTorquato1997,Torquato2002,Torquato2018}. In parallel, a \emph{low-wavenumber} coverage requirement ensures that the smallest Fourier mode of the supercell falls well inside the regime that dictates finite-size bias; we formalize this in \S\ref{subsec:low_k_constraint} (see also the Wiener–Khintchine correspondence recalled in \S\ref{subsec:problem_notation}, Eqs.~\eqref{eq:C}–\eqref{eq:specdens}). 

On the physics side, we cast small-slope capillarity as a linear \emph{screened} (modified Helmholtz/Yukawa) cell problem posed on the periodic torus, with phase-dependent transport in grains and matrix. Periodic homogenization then furnishes an apparent conductivity and an apparent screening parameter, which together define a macroscopic decay length for capillary disturbances; the formulation and cell problems are detailed in \S\ref{subsec:cell_capillarity} and draw on classical capillarity and homogenization theory \cite{DeGennes2004,LandauLifshitz1987,BensoussanLionsPapanicolaou2011,JikovKozlovOleinik1994}, solved efficiently with finite-element (FE) or fast-Fourier-transform (FFT) solver methods \cite{MoulinecSuquet1998,Geers2010}.

Our key observation is that \emph{screened capillarity acts as a spatial low-pass filter}: structural modes with wavenumber below the inverse capillary length contribute most strongly. This motivates a \emph{distribution-aware} treatment of polydispersity. We introduce a capillarity-weighted volume fraction and a \emph{screened} analogue of the integral range that preserve units and reduce to their classical counterparts in the appropriate limits; these are defined precisely in \S\ref{sec:capillarity_distribution_and_sizing}. The resulting sizing rules are quantitative and verifiable: (i) a \emph{length} criterion on the shortest cell edge that depends on a microstructural correlation length, the macroscopic capillary decay length, and a high quantile of grain size (to avoid periodic self-interaction); and (ii) a \emph{volume} criterion that delivers the supercell size needed to meet a target coefficient of variation with a given number of realizations. Directional variants handle anisotropy; a posteriori spectral checks confirm adequate low-$k$ coverage \cite{Torquato2002,Torquato2018}.

In summary, the paper links statistical microstructure descriptors to screened capillary physics and distills them into practical, distribution-aware rules for periodic FE/FFT simulations. It complements reconstruction and descriptor advances \cite{YeongTorquato1998,JiaoPNAS2009} and sharpens RVE-size analyses for random composites \cite{Kanit2003,Ostoja2006} by adding the specific screening length scale and a spectral perspective directly tied to capillarity.

\section{Representative periodic supercell for stationary random, polydisperse granular media}
\label{sec:rve_sizing}

\subsection{Problem statement and notation}
\label{subsec:problem_notation}

We model a \emph{stationary random, polydisperse} granular medium in $\mathbb{R}^3$ with a \emph{fixed (quenched)} microstructure: once a realization is drawn, every grain’s position and size are immutable (no time dependence and no motion). All randomness is defined on a probability space $(\Omega,\mathcal{F},\mathbb{P})$. 
Here $\Omega$ denotes the set of admissible microstructure realizations, 
$\mathcal{F}$ is a $\sigma$–algebra on $\Omega$, and $\mathbb{P}$ is a probability measure on $(\Omega,\mathcal{F})$. 
Ensemble expectations are written $\langle X\rangle := \mathbb{E}[X] = \int_{\Omega} X(\omega)\,d\mathbb{P}(\omega)$.

In particular, we use a \emph{random field indexed by space} (not by time), \hbox{i.e.\ a mapping $\chi:\Omega\times\mathbb{R}^3\to\{0,1\}$} defined below.

A single realization is completely specified by the finite or countable set of grain centers and sizes. To keep the dependence on \emph{position} and \emph{size} explicit, we represent the configuration by a \emph{marked point process} of centers $\mathbf{X}_i\in\mathbb{R}^3$ with positive size marks $A_i\in\mathbb{R}_+$:
\begin{align}
\Phi(\omega) \;=\; \{(\mathbf{X}_i(\omega),A_i(\omega))\}_{i\in\mathbb{N}}
\label{eq:MPP}
\end{align}
which is locally finite in $\mathbb{R}^3\times\mathbb{R}_+$. 

Its associated \emph{position-size random measure} is
\begin{align}
\eta(\mathbf{x},a;\omega) \;:=\; \sum_{i} \delta\!\left(\mathbf{x}-\mathbf{X}_i(\omega)\right)\,\delta\!\left(a-A_i(\omega)\right),
\label{eq:eta}
\end{align}
so the “distribution of granular” is, by construction, a function of position $\mathbf{x}$ and size $a$, with no time argument (grains are fixed). The mean (first moment) is
\begin{align}
f(\mathbf{x},a) \;:=\; \big\langle \eta(\mathbf{x},a;\cdot) \big\rangle .
\label{eq:mean_eta}
\end{align}
By \emph{space-stationarity} (translation invariance in distribution), $f$ has no explicit $\mathbf{x}$-dependence. Under the standard independent-mark assumption (size marks i.i.d.\ and independent of locations),
\begin{subequations}\label{eq:intensity_factorization}
\begin{align}
f(a)&=\nu\,p_A(a),\\
\nu&>0,\qquad \int_0^\infty p_A(a)\,da=1.
\end{align}
\end{subequations}
where $p_A$ is the size density, $\nu$ is the (constant) number density of centers \cite{Chiu2013,DaleyVJ2003}.
The factorization in Eq.~\eqref{eq:intensity_factorization} is a convenient baseline, not a structural requirement of the sizing theory. In dense or jammed packings, steric exclusion and contact networks generally induce correlations between positions and marks, so the independent-mark assumption can fail. The supercell criteria developed below remain applicable provided the relevant two-point descriptors $C(\mathbf r)$, $\widehat C(\mathbf k)$, $A$, or their screened counterparts are estimated directly from the actual microstructure (experimental image, DEM packing, or correlated generator). In that case, Eqs.~\eqref{eq:intensity_factorization} are replaced by the measured statistics of the correlated medium rather than assumed a priori.

From centers and sizes we build the two-phase medium. Let $\mathcal{K}(a)\subset\mathbb{R}^3$ be the grain template of size $a$ (e.g., for spherical grains, $\mathcal{K}(a)$ is the closed ball of radius $a/2$). The granular set in realization $\omega$ is the Boolean (union-of-grains) model
\begin{align}
\Omega_g(\omega) := \bigcup_{i}\big(\mathbf{X}_i(\omega) \oplus \mathcal{K}\big(A_i(\omega)\big)\big).
\label{eq:boolean}
\end{align}
where $\oplus$ denotes the Minkowski sum. The corresponding two-phase indicator random field is
\begin{align}
\chi &: \Omega\times\mathbb{R}^3 \to \{0,1\}, \\[2pt]
&\chi(\omega,\mathbf{x}) \;=\; \mathbf{1}_{\{\mathbf{x}\in\Omega_g(\omega)\}},
\label{eq:chi_def}
\end{align}
i.e., $\chi=1$ inside grains and $\chi=0$ in the matrix \cite{Torquato2002}.

We say the medium is \emph{(strictly) stationary in space} if all finite-dimensional distributions of $\chi$ are invariant under translations:
\begin{equation}
\label{eq:strict_stationarity}
\big(\chi(\cdot,\mathbf{x}_1),\ldots,\chi(\cdot,\mathbf{x}_n)\big)
\stackrel{d}{=}
\big(\chi(\cdot,\mathbf{x}_1+\mathbf{t}),\ldots,\chi(\cdot,\mathbf{x}_n+\mathbf{t})\big).
\end{equation}
This holds  $ \forall n\in\mathbb{N}$, $\{\mathbf{x}_i\}_{i=1}^n\subset\mathbb{R}^3$, $ \forall \mathbf{t}\in\mathbb{R}^3$.

For what follows it suffices to assume \emph{second-order} (wide-sense) stationarity:
\begin{align}
\langle \chi(\mathbf{x}) \rangle &= \phi \in (0,1) \label{eq:mean_stationary}
\end{align}
because $\chi=1$ inside grains, $\phi$ is the solid volume fraction; the porosity is $1-\phi$;
\begin{align}
\mathrm{Cov}\!\big[\chi(\mathbf{x}),\chi(\mathbf{y})\big] &= C(\mathbf{y}-\mathbf{x}). \label{eq:cov_stationary}
\end{align}
The two-point probability function, the autocovariance, and the spatial spectral density (the Fourier transform of $C$) are
\begin{subequations}\label{eq:second_order_stats}
\begin{align}
S_2(\mathbf r) &:= \big\langle \chi(\mathbf x)\,\chi(\mathbf x+\mathbf r)\big\rangle, \label{eq:S2}\\
C(\mathbf r) &:= S_2(\mathbf r)-\phi^2, \label{eq:C}\\
\widehat C(\mathbf k) &:= \int_{\mathbb R^3} C(\mathbf r)\,e^{-i\mathbf k\cdot\mathbf r}\,d\mathbf r. \label{eq:specdens}
\end{align}
\end{subequations}
\noindent
Note that $C$ is even, $C(-\mathbf r)=C(\mathbf r)$, and $C(\mathbf 0)=\phi(1-\phi)$ and the behavior of $\widehat{C}(\mathbf{k})$ as $\mathbf{k}\to\mathbf{0}$ controls the long-wavelength fluctuations that dominate finite-size effects \cite{Torquato2002}.
Moreover, the inverse relation
$C(\mathbf r)=\frac{1}{(2\pi)^3}\int_{\mathbb R^3}\widehat C(\mathbf k)\,e^{i\mathbf k\cdot\mathbf r}\,d\mathbf k$
makes the Fourier pair explicit the Wiener-Khintchine theorem \cite{Wiener1930,Khintchine1934,Yaglom1987}: for stationary \emph{spatial} random fields, the covariance and power spectral density are a Fourier transform pair. In computations we work on a periodic torus (below), where the spectrum is sampled at discrete lattice wavenumbers.

We now introduce the \emph{periodic supercell} (also: periodic cell, representative torus). It is a computational device used to approximate the infinite stationary medium by a periodic extension of a bounded sample. We take an orthorhombic cell and its Bravais lattice:
\begin{align}
Y &= [0,L_x)\times[0,L_y)\times[0,L_z), \\
V &= L_xL_yL_z, \label{eq:cell}\\
\mathbf{L} :&= \mathrm{diag}(L_x,L_y,L_z),\\
\mathcal{L}_{\mathrm{per}} :&= \mathbf L\,\mathbb Z^3, \qquad
\mathbb T^3 = \mathbb R^3/ \mathcal{L}_{\mathrm{per}}. \label{eq:lattice_torus}
\end{align}
Here, we define the torous $\mathbb T^3$ by the discrete additive subgroup (Bravais lattice) $\mathcal{L}_{\mathrm{per}}$.

Given any realization, we restrict it to $Y$ and extend it periodically by wrapping. For any $\mathbf x\in\mathbb R^3$, define the componentwise reduction modulo $Y$ by
\[
[\mathbf x]_Y := \mathbf x - \mathbf L\,\big\lfloor \mathbf L^{-1}\mathbf x \big\rfloor,
\]
where $\lfloor\cdot\rfloor$ acts componentwise. Then set
\begin{align}
\chi_Y(\mathbf{x}) \;&=\; \chi\!\big([\mathbf x]_Y\big), \\ 
\eta_Y(\mathbf{x},a) &=\; \eta\!\big([\mathbf x]_Y,\,a\big),
\label{eq:wrap}
\end{align}
so that $\chi_Y$ and $\eta_Y$ are $\mathcal{L}_{\mathrm{per}}$-periodic. For any integrable $g$ on $Y$, the (torus) average is
\begin{align}
\langle g \rangle_Y \;&:=\; \frac{1}{V}\int_Y g(\mathbf{x})\,d\mathbf{x}.
\label{eq:torus_avg}
\end{align}
Two auxiliary notions used later are the \emph{integral range} (a correlation volume)
\begin{align}
A \;&:=\; \frac{\widehat{C}(\mathbf{0})}{C(\mathbf{0})} \quad \text{(units of volume)},
\label{eq:int_range}
\end{align}
and the smallest nonzero wavenumber resolved by the supercell
\begin{align}
k_{\min} \;&:=\; \frac{2\pi}{\min(L_x,L_y,L_z)} .
\label{eq:kmin}
\end{align}

As a template for subsequent sections, we recall the standard periodic cell problem for a scalar diffusion surrogate. Let
\begin{align}
K(\mathbf{x}) \;=\; K_g\,\chi(\mathbf{x}) \;+\; K_m\,[1-\chi(\mathbf{x})]
\label{eq:K_mix_define}
\end{align}
with $K_g,K_m>0$.  
Both $K_g$ and $K_m$ are the (constant) diffusivities/conductivities in the granular and matrix phases, respectively. For a \emph{uniform macroscopic gradient} $\nabla\bar{u}$ (i.e.\ an affine macroscopic field $\bar{u}(\mathbf{x})=\nabla\bar{u}\cdot\mathbf{x}$), the periodic fluctuation $\tilde u$ solves
\begin{align}
-&\nabla\cdot\!\left(K(\mathbf{x})\big(\nabla\bar{u}+\nabla \tilde{u}\big)\right) \;=\; 0 \quad \text{in } Y, 
\label{eq:cell_pde}\\
&\forall \tilde{u} \in\; H^1_\#(Y), 
\quad \langle \tilde{u}\rangle_Y \;=\; 0,
\label{eq:cell_space}\\[2pt]
&\big\langle K(\mathbf{x})\big(\nabla\bar{u}+\nabla \tilde{u}\big)\big\rangle_Y = K_{\mathrm{app}}(Y)\,\nabla\bar{u},
\label{eq:Kapp}
\end{align}
where $H^1_\#(Y):=\{v\in H^1_{\mathrm{loc}}(\mathbb{R}^3): v \text{ is }Y\text{-periodic}\}$ is the periodic Sobolev space. The zero-mean condition fixes the additive constant. This periodic/affine-periodic setting is Hill--Mandel consistent and is efficiently solvable by FFT- or FE-based methods \cite{Hill1963,Geers2010,MoulinecSuquet1998}. In later sections we use the same periodic framework to treat the \emph{screened} capillarity problem.

\subsection{Integral range and finite-size variance}
\label{subsec:integral_range_variance}

This subsection explains how second-order statistics of the stationary two-phase field $\chi$ determine the supercell size $Y=[0,L_x)\times[0,L_y)\times[0,L_z)$ (volume $V=L_xL_yL_z$) required for representative estimates. The central quantity is the \emph{integral range} (correlation volume)
\begin{align}
A \;=\; \frac{1}{C(\mathbf{0})}\int_{\mathbb{R}^3} C(\mathbf{r})\,d\mathbf{r}
\;=\; \frac{\widehat{C}(\mathbf{0})}{C(\mathbf{0})}
\label{eq:integral_range_repeat}
\end{align}
with $C(\mathbf 0)=\phi(1-\phi),\quad A>0,$ which compresses the overall strength of spatial correlations into a single volume scale (typically $A=O(\xi^3)$ when $C$ decays beyond a correlation length $\xi$) \cite{Torquato2002}.

We begin with the periodic estimator of the volume fraction,
\begin{align}
\widehat{\phi}(Y)\;=\;\frac{1}{V}\int_Y \chi(\mathbf{x})\,d\mathbf{x}.
\label{eq:phihat}
\end{align}
Using stationarity and the change of variables $(\mathbf x,\mathbf y)\mapsto(\mathbf x,\mathbf r=\mathbf y-\mathbf x)$ yields the exact \emph{window identity} \cite{LuTorquato1990,QuintanillaTorquato1997,Torquato2002}:
\begin{align}
\operatorname{Var}\!\big[\widehat{\phi}(Y)\big]
&= \frac{1}{V^2}\int_Y\!\!\int_Y C(\mathbf{y}-\mathbf{x})\,d\mathbf{x}\,d\mathbf{y} \nonumber \\
&= \frac{1}{V}\int_{\mathbb{R}^3} C(\mathbf{r})\,\alpha_Y(\mathbf{r})\,d\mathbf{r},
\label{eq:window_identity}
\end{align}
where $\alpha_Y(\mathbf r):=\frac{1}{V}\int_{\mathbb{R}^3}\mathbf 1_Y(\mathbf x)\,\mathbf 1_Y(\mathbf x+\mathbf r)\,d\mathbf x$ is the normalized intersection volume of $Y$ with its translate. For a rectangular box,
\begin{align}
\alpha_Y(\mathbf r)=\prod_{\alpha\in\{x,y,z\}}\Big(1-\frac{|r_\alpha|}{L_\alpha}\Big)_+,
\quad (u)_+:=\max\{u,0\}.
\label{eq:alpha_rect}
\end{align}
If $C\in L^1(\mathbb R^3)$ (short-range correlations), then $\alpha_Y(\mathbf r)\to 1$ for fixed $\mathbf r$ as $\min(L_x,L_y,L_z)\to\infty$, and the leading asymptotics is
\begin{align}
\operatorname{Var}\!\big[\widehat{\phi}(Y)\big]
&= \frac{1}{V}\int_{\mathbb R^3} C(\mathbf r)\,d\mathbf r\ +\ o\!\Big(\frac{1}{V}\Big) \nonumber \\
&= \frac{C(\mathbf0)\,A}{V}\ +\ o\!\Big(\frac{1}{V}\Big) \nonumber \\
&= \frac{\phi(1-\phi)\,A}{V}\ +\ o\!\Big(\frac{1}{V}\Big),
\label{eq:var_phi}
\end{align}
i.e., the classical $1/V$ decay of local volume-fraction variance \cite{LuTorquato1990,QuintanillaTorquato1997,Torquato2002}. The same result appears in Fourier space by writing $W_Y(\mathbf k):=\int_Y e^{-i\mathbf k\cdot\mathbf x}\,d\mathbf x$ and using
\begin{align}
\operatorname{Var}\!\big[\widehat{\phi}(Y)\big]
= \frac{1}{(2\pi)^3\,V^2}\int_{\mathbb R^3} |W_Y(\mathbf k)|^2\,\widehat C(\mathbf k)\,d\mathbf k,
\label{eq:spectral_var}
\end{align}
where the kernel $|W_Y|^2/V^2$ concentrates near $\mathbf k=\mathbf 0$ as $Y$ grows, so the leading term is governed by $\widehat C(\mathbf 0)=\int C$ and hence by $A$ \cite{LuTorquato1990,Torquato2002}.

Beyond volume fraction, many scalar apparent properties $P(Y)$ computed from periodic cell problems (e.g., a component of effective conductivity, stiffness, or the screening coefficient in capillarity) exhibit an analogous scaling:
\begin{align}
\operatorname{Var}\!\big[P(Y)\big]\;\approx\;\frac{C_P\,A}{V},
\label{eq:var_P}
\end{align}
with a finite prefactor $C_P=O(1)$ that depends on the physics (contrast, loading, boundary conditions) but not on $V$ \cite{Kanit2003,Ostoja2006}. 
In \eqref{eq:var_P}, $V=L_xL_yL_z$ is the supercell volume; 
$A$ is the integral range (Eq.~\eqref{eq:int_range}, units of volume), which measures the correlation volume of $\chi$;
and $C_P$ is a physics-dependent prefactor with the units of $P^2$ (e.g., $C_P=C(\mathbf 0)=\phi(1-\phi)$ for $P=\widehat\phi$, and Eq.~\eqref{eq:var_beta_explicit} for $P=\beta_{\rm app}$). 
Thus $\operatorname{Var}[P(Y)]$ decays like (correlation volume)/(sample volume), scaled by the property sensitivity.
Averaging $n$ independent realizations gives
\begin{align}
\operatorname{Var}\!\Big[\overline{P}_n(Y)\Big]
= \frac{1}{n}\,\operatorname{Var}\!\big[P(Y)\big]
\;\approx\; \frac{C_P\,A}{n\,V},
\label{eq:var_avg}
\end{align}
and the resulting sizing rule for a target coefficient of variation $\mathrm{CV}\le\varepsilon$ is
\begin{align}
V_{\min}\;\gtrsim\;\frac{C_P\,A}{n\,\varepsilon^2}.
\label{eq:cv_rule}
\end{align}

To make $C_P$ concrete for \emph{diffusion}, fix a unit macroscopic loading $\mathbf{E}\in\mathbb{R}^3$ and consider the scalar response
\begin{align}
\label{eq:P_scalar_def}
P(Y;\mathbf{E}) :&= \mathbf{E}^{\top} K_{\mathrm{app}}(Y)\,\mathbf{E}
= E_i\, (K_{\mathrm{app}})_{ij}(Y)\, E_j \nonumber \\
& = K_{\mathrm{app}}(Y):\big(\mathbf{E}\otimes\mathbf{E}\big).
\end{align}
At first order, the variance prefactor can be expressed as a covariance-weighted integral of a \emph{sensitivity density} $s(\mathbf x)$:
\begin{align}
\operatorname{Var}[P(Y)]
&\approx \frac{1}{V}\int_{\mathbb{R}^3} \Gamma(\mathbf{r})\,C(\mathbf{r})\,d\mathbf{r},
\label{eq:CP_general_fixed}
\\
C_P
&= \frac{1}{A}\int_{\mathbb{R}^3} \Gamma(\mathbf{r})\,C(\mathbf{r})\,d\mathbf{r}.
\label{eq:CP_formula_fixed}
\end{align}
where $\Gamma(\mathbf{r})=\lim_{Y\uparrow\mathbb{R}^3}\frac{1}{V}\int_Y s(\mathbf{x})\,s(\mathbf{x}+\mathbf{r})\,d\mathbf{x}$ is the autocorrelation of $s$ in the stationary limit. Therefore,
\begin{align}
C_P \;=\; \frac{1}{C(\mathbf{0})}\int_{\mathbb{R}^3} \Gamma(\mathbf{r})\,C(\mathbf{r})\,d\mathbf{r},
\label{eq:CP_formula}
\end{align}
which shows explicitly that, for diffusion, $C_P$ depends on the phase conductivities $(K_g,K_m)$ (through $s$), on the volume fraction, and on the loading.

Assume the apparent property $P(Y)$ depends predominantly on the local
volume fraction measured in the cell, i.e., $P(Y)\approx g(\widehat{\phi}(Y))$
for a smooth scalar map $g$.
A first-order Taylor (delta-method) expansion about the ensemble mean
$\phi=\langle\widehat{\phi}\rangle$ gives
\begin{equation}
P(Y)\;\approx\; g(\phi)\;+\;g'(\phi)\,\big(\widehat{\phi}(Y)-\phi\big).
\label{eq:delta_expansion}
\end{equation}
Taking the variance and neglecting higher-order terms,
\begin{equation}
\operatorname{Var}[P(Y)]\;\approx\;\big(g'(\phi)\big)^{2}\,
\operatorname{Var}\!\big[\widehat{\phi}(Y)\big].
\label{eq:delta_variance}
\end{equation}
Using \eqref{eq:var_phi} for $\operatorname{Var}[\widehat{\phi}(Y)]$ yields
\begin{equation}
\operatorname{Var}[P(Y)]
\;\approx\; \frac{A}{V}\,\phi(1-\phi)\,\big(g'(\phi)\big)^2,
\label{eq:CP_volfrac}
\end{equation}
which is the form used below.

In dilute/weak-contrast \emph{isotropic} media where the effective scalar conductivity depends primarily on $\phi$, set $g(\phi)=K_{\mathrm{eff}}^{\mathrm{Max}}(\phi)$ (Maxwell’s dilute-sphere estimate),
\begin{align}
K_{\mathrm{eff}}^{\mathrm{Max}}(\phi)
=
K_m\,\frac{K_g+2K_m+2\phi\,(K_g-K_m)}{K_g+2K_m-\phi\,(K_g-K_m)},
\label{eq:Maxwell}
\end{align}
so that
\begin{align}
\frac{dP}{d\phi}
\;=\;
\frac{3\,K_m\,(K_g-K_m)\,(K_g+2K_m)}
{\big(K_g+2K_m-\phi\,(K_g-K_m)\big)^{2}}.
\label{eq:dMaxwell}
\end{align}
Using \eqref{eq:var_phi} in \eqref{eq:CP_volfrac} gives the explicit diffusion prefactor
\begin{align}
C_P^{\mathrm{diff}}
\approx
\phi(1-\phi)
\left[\frac{3\,K_m\,(K_g-K_m)\,(K_g+2K_m)}
{\big(K_g+2K_m-\phi\,(K_g-K_m)\big)^{2}}\right]^{\!2},
\label{eq:CP_explicit}
\end{align}
which makes the dependence on $(K_g,K_m,\phi)$ fully transparent. For stronger contrasts or non-dilute regimes, one may replace \eqref{eq:Maxwell} by a self-consistent estimate (e.g., Bruggeman) or compute $dP/d\phi$ numerically; the scaling $\operatorname{Var}[P(Y)]\approx (C_P A)/V$ remains the guiding law \cite{Kanit2003,Ostoja2006}.

Two remarks delimit the regime of validity. If $C\notin L^1(\mathbb R^3)$ (very long-range correlations), then $A$ diverges and the $1/V$ law breaks down; variance decays more slowly and much larger cells are required \cite{Torquato2002}. In \emph{hyperuniform} media one has $\widehat C(\mathbf 0)=\int C=0$, hence $A=0$ and the variance decays \emph{faster} than $1/V$, with exponents set by the small-$k$ behavior of $\widehat C$ \cite{Torquato2018}.

\subsection{Low-wavenumber resolution (length) constraint}
\label{subsec:low_k_constraint}

On the periodic cell $Y=[0,L_x)\times[0,L_y)\times[0,L_z)$ (volume $V=L_xL_yL_z$), any $\mathcal{L}_{\mathrm{per}}$-periodic (i.e., $Y$-periodic) field admits a \emph{discrete} spatial Fourier representation. The admissible wavevectors form the \emph{wavenumber lattice}
\begin{align}
\mathcal{K}(Y)
\;=\;
\Big\{
\mathbf{k}(\mathbf{n}&)
=
\Big(\tfrac{2\pi n_x}{L_x},\tfrac{2\pi n_y}{L_y},\tfrac{2\pi n_z}{L_z}\Big)
 \nonumber \\
& : \mathbf{n}=(n_x,n_y,n_z)\in\mathbb{Z}^3
\Big\}.
\label{eq:k_lattice}
\end{align}
This discreteness is a consequence of the \emph{finite periodic domain} (i.e., the torus $\mathbb{T}^3=\mathbb{R}^3/\mathcal{L}_{\mathrm{per}}$), not of any numerical mesh used later to discretize PDEs. Two low-$k$ scales are useful:

\begin{align}
k_{\min}^{\mathrm{mag}}
&:=\min_{\mathbf{n}\neq\mathbf{0}}\big|\mathbf{k}(\mathbf{n})\big|
\;=\; \frac{2\pi}{\max(L_x,L_y,L_z)},
\label{eq:kmin_mag}\\[4pt]
k_{\min}
&:= \frac{2\pi}{\min(L_x,L_y,L_z)}.
\label{eq:kmin_iso}
\end{align}

The first is the smallest \emph{magnitude} present (found by placing $\pm1$ along the longest edge and zeros otherwise). The second is a conservative \emph{directional} resolution ensuring access to small wavenumbers \emph{along every axis}. In what follows we adopt \eqref{eq:kmin_iso} because it guarantees low-$k$ coverage irrespective of aspect ratio.

The finite-size variance of periodic estimators (e.g., \eqref{eq:var_phi}) is governed by the \emph{low-wavenumber} content of the spectral density $\widehat C(\mathbf{k})$ (cf.\ \eqref{eq:spectral_var}). To sample that regime faithfully we require that the smallest resolved wavenumber lies well below the crossover scale $k_c$ where $\widehat C(\mathbf{k})$ leaves its low-$k$ plateau/slope:
\begin{align}
k_{\min}\ \ll\ k_c
\Longleftrightarrow
\min(L_x,L_y,L_z)\ \gg\ \frac{2\pi}{k_c}.
\label{eq:lowk_criterion}
\end{align}
If the shortest edge is too small, the window spectrum $|W_Y(\mathbf{k})|^2/V^2$ never probes the asymptotic low-$k$ region of $\widehat C$, and the $A/V$ variance law (Section~\ref{subsec:integral_range_variance}) is not yet observed \cite{Torquato2002}.

In practice $k_c$ is estimated, not known. A robust surrogate uses a correlation length $\xi$ extracted from $C(\mathbf r)$ (e.g., $1/e$ decay or second-moment correlation length, when finite) or equivalently from the low-$k$ width of $\widehat C$. This leads to the working rule
\begin{align}
\min(L_x,L_y,L_z)\ \gtrsim\ c\,\xi,
\qquad c=\mathcal{O}(10).
\label{eq:L_vs_xi}
\end{align}
Here $c=\mathcal{O}(10)$ is a \emph{heuristic safety factor}, not a universal constant: values $c\in[8,12]$ typically place $k_{\min}$ at least a factor $5$–$10$ below the knee of $\widehat C(k)$ for short-ranged covariances, which in turn makes the normalized intersection factor $\alpha_Y(\mathbf r)$ close to $1$ over the main support of $C$. The appropriate $c$ depends on how sharply $\widehat C(k)$ turns near $k=0$ and on the acceptable bias. We therefore \emph{verify a posteriori} by plotting (optionally isotropized) $\widehat C(k)$, marking $k_{\min}=2\pi/\min(L_\alpha)$, and confirming $k_{\min}\ll k_c$ for the material at hand \cite{Torquato2002,Torquato2018}. For hyperuniform media, where $\widehat C(\mathbf 0)=0$ and $\widehat C(\mathbf k)\sim |\mathbf k|^\alpha$ as $\mathbf k\to \mathbf 0$, $L$ must be chosen so that $k_{\min}$ lies inside the power-law regime; this can require $L$ beyond the conservative $c\,\xi$ threshold \cite{Torquato2018}.

The numerical constants $c_\xi$ and $c_\lambda$ in Eqs.~\eqref{eq:L_vs_xi} and \eqref{eq:size_cap} are deliberately specified as order--of--magnitude factors, $c_\xi,c_\lambda=\mathcal{O}(10)$, rather than universal thresholds. 
In practice we choose $c_\xi$ by inspecting $\widehat C(k)$: $c_\xi\simeq 8$--$12$ places $k_{\min}=2\pi/\min(L_\alpha)$ one decade or more below the spectral knee $k_c$ for short-ranged covariances. 
If $c_\xi$ or $c_\lambda$ are underestimated by, say, $20\%$, then $k_{\min}$ is shifted closer to the knee; the numerical example in Sec.~\ref{sec:numerical_validation} shows that this primarily manifests as a modest upward deviation from the asymptotic $A/V$ variance line for the smallest boxes, rather than a catastrophic breakdown. 
Overestimates of the same magnitude simply yield \emph{conservative} (over-sized) boxes: the spectral coverage is more than sufficient and the variance still follows the predicted $1/V$ law, at the expense of additional computational cost. 
Thus the prefactors control a precision--cost trade-off, but the structure of the sizing rules and the underlying $A/V$ scaling are robust to moderate variations.
Two practical refinements are helpful. First, control the aspect ratio to avoid missing low–$k$ along a short axis:
\begin{align}
\frac{\min(L_x,L_y,L_z)}{\max(L_x,L_y,L_z)} \;\ge\; \rho_{\min},
\label{eq:aspect}
\end{align}
with $\rho_{\min}\in(0,1) \ \ \text{(e.g., }\rho_{\min}\approx\tfrac{1}{2}\text{)},$ so that the shortest edge is not excessively smaller than the longest.

Second, distinguish low–$k$ reach (set by $L_\alpha$) from numerical resolution (set by mesh spacing $\Delta x_\alpha$). For a uniform grid, the highest resolvable wavenumber along direction $\alpha$ is the Nyquist cutoff
\begin{align}
k_{\mathrm{Nyq},\alpha} \;=\; \frac{\pi}{\Delta x_\alpha},
\label{eq:grid_note}
\end{align}
which affects high–$k$ accuracy but does not change the low–$k$ reach determined by $k_{\min}=2\pi/\min(L_\alpha)$ in \eqref{eq:kmin_iso}. Choose $L_\alpha$ to satisfy the low–$k$ criteria, and choose $\Delta x_\alpha$ for discretization accuracy.

Finally, recall that the \emph{volume} requirement $V\gtrsim (C_P A)/(n\varepsilon^2)$ in \eqref{eq:cv_rule} comes from Section~\ref{subsec:integral_range_variance}: it ensures the desired statistical precision, where $n$ is the number of independent realizations and $\varepsilon$ the target coefficient of variation. The present length constraint \eqref{eq:lowk_criterion}–\eqref{eq:L_vs_xi} is its \emph{spectral complement}: the former fixes \emph{how much} material to average, while the latter ensures we average over the \emph{right wavelengths}.

\subsection{Geometric safeguard for polydispersity}
\label{subsec:geom_safeguard}

Polydispersity imposes a simple but important \emph{geometric} constraint on the shortest edge of the periodic cell. Let $D$ denote the (random) grain diameter (the size mark from \S\ref{subsec:problem_notation}), with PDF $p_D(a)$. Choose a high, but not extreme, upper quantile to represent the “largest relevant” grains:
\begin{align}
a_{\max}
\;:=\;
Q_{1-\delta}(D)
\;=\;
\inf\{\,a:\ \mathbb{P}(D\le a)\ge 1-\delta\,\},
\label{eq:amax_def}
\end{align}
with $\delta\in[10^{-3},10^{-2}] \ \text{(typ.)}$.
Using a quantile avoids letting rare outliers dictate the box size while still accounting for the tail of the size distribution.

A first baseline is purely geometric. Place a spherical grain of diameter $a$ with its center on a periodic face; its nearest periodic image is displaced by the \emph{shortest} box vector $\min(L_x,L_y,L_z)$. To prevent \emph{self-overlap} across that face, the center separation must be at least $a$:
\begin{align}
\min(L_x,L_y,L_z)\ \ge\ a_{\max}
\label{eq:no_overlap}
\end{align}
with no geometric self-overlap of the largest grains. 
Condition \eqref{eq:no_overlap} prevents literal overlap but not \emph{strong near-field self-interactions} (near contacts, lubrication/contact forces, or neighbor-list kernels) between a grain and its periodic images when the box is only one diameter thick.

A second, widely used principle is the \emph{minimum-image convention} for short-range interactions with cut-off $R_c$: to ensure that only one image of any particle contributes within the cut-off, one requires
\begin{align}
R_c \;<\; \tfrac{1}{2}\,\min(L_x,L_y,L_z),
\label{eq:mic_rule}
\end{align}
as stated in standard Molecular Dynamics texts and manuals \cite{AllenTildesley2017,GromacsManual}. If the relevant interaction range scales with the large-grain size, say $R_c\simeq \gamma\,a_{\max}$ (with $\gamma\in[1,2]$ depending on the physics—hard/soft contact shells, lubrication, neighbor-search radius), then \eqref{eq:mic_rule} implies
\begin{align}
\min(L_x,L_y,L_z)\ \gtrsim\ 2\,\gamma\,a_{\max}
\;\;\in\;\;[2,4]\;a_{\max}.
\label{eq:mic_implied}
\end{align}
To suppress spurious near-field correlations between periodic images of the \emph{largest} grains (especially in highly polydisperse media and for non-spherical shapes, where a circumscribed diameter is more appropriate), we adopt a conservative buffer:
\begin{align}
\min(L_x,L_y,L_z)\ \gtrsim\ c\,a_{\max},
\qquad c\in[3,5],
\label{eq:geom_guard}
\end{align}
which strictly contains both the non-overlap condition \eqref{eq:no_overlap} and the minimum-image implication \eqref{eq:mic_implied} for typical $\gamma$. For markedly anisotropic grains (fibers, plates, elongated clasts), $a_{\max}$ should be interpreted as the upper quantile of a \emph{geometric extent relevant to periodic self-interaction}, not as an equivalent-sphere diameter. A robust choice is the maximum chord length, or equivalently the diameter of the circumscribed sphere, for the grain family:
\[
a_{\max}^{\rm geom}:=Q_{1-\delta}(d_{\rm circ}).
\]
For highly elongated particles, near-parallel image interactions can persist over distances larger than the minor axis, so the safety factor $c_a$ may need to be chosen toward the upper end of, or modestly above, the range $[3,5]$. If the medium is directionally anisotropic, a directional safeguard based on projected extents along each axis may be used instead.

The safeguard \eqref{eq:geom_guard} is orthogonal to the variance and low-$k$ constraints from \S\ref{subsec:integral_range_variance}–\S\ref{subsec:low_k_constraint}. The $A/V$ law sets \emph{how much} volume is needed for statistical stabilization \cite{Kanit2003,Ostoja2006,Torquato2002}, and $k_{\min}\ll k_c$ ensures the box is \emph{long enough} to probe the correct small-$k$ regime \cite{Torquato2002,Torquato2018}. By contrast, \eqref{eq:geom_guard} is a local, geometry-driven hygiene condition that prevents the few largest grains from interacting too strongly with their own periodic copies. Enforcing \eqref{eq:geom_guard} avoids artifacts (image self-contact, near-contact bias, aliasing of neighbor shells) that can persist even if $V$ is large and $k_{\min}$ is small. In the same spirit, the choice $c_a\in[3,5]$ in Eq.~\eqref{eq:geom_guard} should be viewed as a conservative buffer: choosing $c_a$ too small can leave residual self-interaction of the largest grains (as seen in Sec.~\ref{subsec:self_interaction_numerics}), whereas choosing it slightly larger only increases the box volume without altering the form of the sizing rules.

It is cheap to evaluate from the size distribution alone via \eqref{eq:amax_def}, requires no PDE solves or spectrum estimates, and provides a quick sanity check before expensive sampling. It also matches the minimum-image requirement \eqref{eq:mic_rule} standard in particle simulations with periodic boundaries \cite{AllenTildesley2017,GromacsManual}. In practice, apply \eqref{eq:geom_guard} first (cheap), then verify statistical targets with \eqref{eq:cv_rule} and spectral reach with \eqref{eq:lowk_criterion}.

 \subsection{Screened Darcy with Phase-Dependent Transport }

For small interfacial slopes under gravity, the linearized Young--Laplace balance
$\gamma\,\nabla^2 h = \Delta\rho\,g\,h$ implies a screened, modified-Helmholtz operator for the interface height,
$\nabla^2 h - \lambda^{-2} h = 0$ with capillary length $\lambda^2=\gamma/(\Delta\rho g)$. The associated Green function in 2D decays as $K_0(r/\lambda)$ (where \( K_0 \) is the modified Bessel function of the second kind), highlighting exponential screening.  These results are classical in capillarity, see e.g. \cite{Kralchevsky2000,DeGennes2004}. In porous media, Darcy-type models employ phase-dependent transport (mobilities/relative permeabilities), often cast via a global-pressure formulation and capillary pressure closures \cite{Leverett1941,ChaventJaffre1986}. Together, these strands motivate a bulk equation with a Darcy operator augmented by a capillary screening term and phase-dependent transport coefficients.

Let $p$ denote pressure ($p=\Delta\rho g h$, where $h$ is hydraulic head) and let $K(\chi)$ be a transport coefficient depending on a phase descriptor $\chi$ (e.g., granular, wetting or not wetting fluid). A screened Darcy balance reads
\begin{equation}
\label{eq:screened-darcy}
-\nabla\cdot\big(K(\chi)\nabla p\big)+K(\chi)\lambda^{-2}p= \rho_0
\end{equation}
where $\lambda=\sqrt{\gamma/(\Delta\rho g)}$ encodes capillarity (surface tension $\gamma$, density contrast $\Delta\rho$, gravity $g$), and $\rho_0$ collects injections/ejections. 
In the limit $\lambda\to\infty$, Eq.~\eqref{eq:screened-darcy} reduces to the classical Darcy/Poisson form. With constant $K$, Eq.~\eqref{eq:screened-darcy} is the modified Helmholtz (Yukawa) equation. Numerical aspects of the modified Helmholtz operator and its Green-function structure are discussed in \cite{Winkelmann2021}.

We stress that this screened formulation targets the \emph{macroscopic capillary field} in the small-slope regime. It is not intended to resolve strongly nonlinear pore-scale events such as snap-off, burst dynamics, or Haines jumps, whose onset depends on local throat geometry, network connectivity, and dynamic contact-line effects beyond the present linearized homogenized description. In particular, flowrate-controlled invasion percolation \cite{WilkinsonWillemsen1983} and pressure-controlled invasion with trapping \cite{AlonsoAndersson2025PRE} exhibit scale-free (power-law) cluster statistics and long-range connectivity, which can shift REV sizing towards invasion correlation lengths and finite-size scaling rather than a finite correlation volume alone. In practice, when invasion criticality is the target observable (e.g., cluster-size or burst-size exponents, trapped-phase topology), REV sizing should instead be based on the corresponding correlation length (or finite-size scaling) of the invasion process, which can grow to the system size near threshold.

\subsection{Periodic cell problem for capillarity (screened/modified Helmholtz)}
\label{subsec:cell_capillarity}

We consider a periodic computational cell $Y=[0,L_x)\times[0,L_y)\times[0,L_z)$ that is tiled to form the 3D torus. Within this cell the granular microstructure enters through the phase indicator $\chi(\mathbf x)\in\{0,1\}$ (equal to 1 in grains and 0 in matrix). In the small-slope limit of the Young--Laplace law under gravity, the height field $h(\mathbf x)$ of a fluid interface above a reference plane satisfies a linear balance between interfacial curvature and hydrostatics. Eliminating pressure using hydrostatic balance yields the classical \emph{capillary length}
\begin{align}
\lambda^2 \;=\; \frac{\gamma_{\mathrm{nw}}}{(\rho_{\mathrm{w}}-\rho_{\mathrm{nw}})\,g},
\qquad
\Delta\rho := \rho_{\mathrm{w}}-\rho_{\mathrm{nw}},
\label{eq:lambda_def_cap}
\end{align}
and the linearized equilibrium equation takes the \emph{modified (screened) Helmholtz} form $\nabla^2 h - \lambda^{-2} h = 0$ in homogeneous media \cite{Kralchevsky2000,Vella2005}.
Here, $\rho_{\mathrm{nw}}$ and $\rho_{\mathrm{w}}$ denote the nonwetting and wetting fluid densities, respectively, and $\gamma_{\mathrm{nw}}$ is the interfacial tension of that fluid pair. Thus $\lambda$ is set by the fluid–fluid interface that resides \emph{in the pore space}.
The screening parameters $(\gamma_{\mathrm{nw}},\rho_{\mathrm{w}}-\rho_{\mathrm{nw}})$ pertain to the wetting/nonwetting fluid pair occupying the \emph{pore} space; the solid skeleton only enters through the geometry (which sets the pore connectivity and $K_{\mathrm{p}}$).

In a heterogeneous two-phase medium we allow the transport coefficient to vary with phase. 
Thus, in the capillarity subsections we need transport only in the pore network. We therefore interpret the two-phase law as a \emph{pore/solid} mixture with
\[
K(\mathbf x)=K_{\mathrm{s}}\,\chi(\mathbf x)+K_{\mathrm{p}}\,[1-\chi(\mathbf x)],
\]
where $\chi=1$ in \emph{solids} and $1-\chi$ in \emph{pores} (porosity $\phi_{\mathrm{p}}:=1-\phi$). Here $K_{\mathrm{p}}$ is an \emph{effective} pore mobility (e.g., a Darcy mobility depending on mean saturation), whereas $K_{\mathrm{s}}\ll K_{\mathrm{p}}$ is a small stabilizing value in the solid (often taken $\to 0$ in practice). This keeps ellipticity on the torus while restricting transport to the pore space.
We model the (scalar) conductivity as
\begin{align}
K(\mathbf x) \;=\; K_g\,\chi(\mathbf x) \;+\; K_m\,\big[1-\chi(\mathbf x)\big],
\label{eq:K_mix_cap}
\end{align}
with $K_g>0, \, K_m>0$, where $K_g$ and $K_m$ are, respectively, the grain and matrix coefficients (hydraulic conductivities if $h$ is hydraulic head, or mobilities if one works with pressure). 
In the capillarity context we read $(K_g,K_m)\equiv (K_{\mathrm{s}},K_{\mathrm{p}})$ and $\;1-\phi=\phi_{\mathrm{p}}$ (porosity).
Grouping gravity and surface tension into a local screening coefficient leads to
\begin{align}
c(\mathbf x) \;:=\; \frac{K(\mathbf x)}{\lambda^2}.
\label{eq:c_local}
\end{align}
The steady periodic boundary-value problem on $Y$ is then
\begin{align}
-\nabla\!\cdot\!\big(K(\mathbf x)\,\nabla h(\mathbf x)\big)\;+\;c(\mathbf x)\,h(\mathbf x) \;=\; \rho_0(\mathbf x)
\quad \text{in } Y,
\label{eq:screened_strong_cap}
\end{align}
with $h$ is $Y$-periodic, which is a \emph{screened Poisson} (a.k.a.\ modified Helmholtz or Yukawa) problem with periodic boundary conditions. The diffusion term spreads $h$, the reaction term relaxes it over the length $\lambda$, and $\rho_0$ represents sources/sinks. Because $c(\mathbf x)\ge 0$ and $K(\mathbf x)>0$ a.e., the reaction term removes the constant-function kernel that appears in pure diffusion; hence no compatibility condition on $\rho_0$ is required. This screened structure and its free-space Green functions are standard \cite{Winkelmann2021}.

For computation we use the weak form: given any $Y$-periodic $v\in H^1(Y)$,
\begin{align}
&\int_Y K(\mathbf x)\,\nabla h \!\cdot\! \nabla v\,d\mathbf x 
\;+\; \nonumber \\
&\int_Y c(\mathbf x)\,h\,v\,d\mathbf x
\;=\;
\int_Y \rho_0(\mathbf x)\,v(\mathbf x)\,d\mathbf x.
\label{eq:screened_weak_cap}
\end{align}
Periodic finite elements assemble \eqref{eq:screened_weak_cap} from standard $C^0$ bases, while FFT-based solvers leverage periodicity by treating the diffusion part in Fourier space and applying the reaction term pointwise in real space \cite{MoulinecSuquet1998}. A quick units check in the head formulation confirms consistency:
\begin{align}
&\nabla\!\cdot(K\nabla h):\ [L/T]\cdot(1/L)\to[1/T] \nonumber \\
&c\,h:\ \big([L/T]/L^2\big)\cdot[L]\to[1/T].
\label{eq:units_check}
\end{align}

To pass from the microproblem to a macromodel, we identify an apparent (homogenized) description
\begin{align}
-\nabla\!\cdot\!\big(K_{\rm app}\,\nabla H\big) \;+\; \beta_{\rm app}\,H \;=\; \overline{\rho}_0,
\label{eq:macro_screened}
\end{align}
where $K_{\rm app}$ is the effective conductivity tensor and $\beta_{\rm app}$ the effective screening. The tensor $K_{\rm app}$ follows from the usual gradient-loading correctors: for each Cartesian direction $\alpha\in\{x,y,z\}$, find a $Y$-periodic, zero-mean corrector $w^{(\alpha)}$ such that
\begin{align}
&\int_Y K(\mathbf x)\,\big(\mathbf e_\alpha+\nabla w^{(\alpha)}\big)\!\cdot\!\nabla v\,d\mathbf x = 0
\quad \forall v\in H^1(Y) \\
&K_{\rm app}\,\mathbf e_\alpha \;=\; \Big\langle K(\mathbf x)\,\big(\mathbf e_\alpha+\nabla w^{(\alpha)}\big)\Big\rangle_Y.
\label{eq:K_app_from_cell}
\end{align}
Because a unit macroscopic gradient has no zero-wavenumber content, the reaction term does not enter \eqref{eq:K_app_from_cell} at first order. The apparent screening is obtained either by a single \emph{uniform-source} solve,
\begin{align}
-\nabla\!\cdot\!\big(K\,\nabla \psi\big)\;+\;c\,\psi &= 1 \ \text{ in }Y, \\
\beta_{\rm app} &= \frac{1}{\langle \psi\rangle_Y},
\label{eq:beta_via_mean}
\end{align}
with $\psi$ periodic or directly from first-order periodic homogenization of reaction--diffusion, which yields the simple average
\begin{align}
\beta_{\rm app} \;=\; \langle c\rangle_Y \;=\; \Big\langle \frac{K(\mathbf x)}{\lambda^2}\Big\rangle_Y,
\label{eq:beta_average}
\end{align}
see, e.g., classical results in \cite{BensoussanLionsPapanicolaou2011,JikovKozlovOleinik1994}. For an isotropic scalar summary one may define an \emph{apparent capillary length}
\begin{align}
\frac{1}{\lambda_{\rm app}^2} &:= \frac{\beta_{\rm app}}{K_{\rm iso}}, \\
K_{\rm iso} &:= \tfrac{1}{3}\,\mathrm{tr}\,K_{\rm app}, \\
\lambda_{\rm app} &= \sqrt{\frac{K_{\rm iso}}{\beta_{\rm app}}}.
\label{eq:lambda_app_scalar}
\end{align}

This formulation is useful for two reasons. First, it provides a numerically robust periodic cell problem that directly returns $K_{\rm app}$ and $\beta_{\rm app}$ from the microstructure, giving a compact macro model \eqref{eq:macro_screened}. Second, the effective decay length $\lambda_{\rm app}$ supplies a \emph{physics-based} length scale for sizing the periodic supercell in capillarity-dominated problems: the shortest cell edge should be several $\lambda_{\rm app}$ to resolve the long-wavelength (low-$k$) regime of the screened response. Finally, there is a natural connection to pore-morphology approaches: recent work by Alonso--Marroqu\'{\i}n and Andersson relates capillary pressure--saturation curves to morphology with minimal physics inputs (contact angle and interfacial tension) \cite{AlonsoAndersson2025PRE}. While their focus is quasistatic drainage/imbibition, our $\lambda_{\rm app}$ plays a complementary role as the dynamic screening length of capillary disturbances in continuum models; it can be compared with pore-throat statistics that control entry pressures in morphology-based analyses.

\subsection{Choosing the supercell size for capillarity}
\label{subsec:capillarity_sizing}

The periodic cell $Y=[0,L_x)\times[0,L_y)\times[0,L_z)$ must be large enough in four complementary senses. 
First, it must resolve the material’s long-wavelength statistics. Let $\xi$ denote a correlation length inferred from $C(\mathbf r)$ or from the low-$k$ width of $\widehat C(\mathbf k)$. Requiring the window spectrum $|W_Y|^2/V^2$ to reach the small-$k$ regime that controls finite-size bias leads to
\begin{align}
\min(L_x,L_y,L_z)\ \gtrsim\ c_\xi\,\xi,
\qquad c_\xi=\mathcal{O}(10),
\label{eq:size_micro}
\end{align}
which is the practical form of the low-$k$ criterion in \S\ref{subsec:low_k_constraint} and ensures the $A/V$ variance scaling of \S\ref{subsec:integral_range_variance}.

Second, it must resolve the \emph{screened} capillary response. After solving the periodic capillarity cell problems of \S\ref{subsec:cell_capillarity}, we obtain $K_{\rm app}$ and $\beta_{\rm app}$, and define
\begin{align}
\frac{1}{\lambda_{\rm app}^2} &= \frac{\beta_{\rm app}}{K_{\rm iso}}, \\
K_{\rm iso}&:=\tfrac{1}{3}\,\mathrm{tr}\,K_{\rm app}.
\label{eq:lambda_app_repeat}
\end{align}
To place the smallest wavenumber $k_{\min}=2\pi/\min(L_\alpha)$ well inside the low-$k$ band of the screened operator, we require
\begin{align}
\min(L_x,L_y,L_z)\ \gtrsim\ c_\lambda\,\lambda_{\rm app},
\qquad c_\lambda=\mathcal{O}(10),
\label{eq:size_cap}
\end{align}
so that $k_{\min}\ll 1/\lambda_{\rm app}$. For anisotropic $K_{\rm app}$, a directional variant is
\begin{subequations}\label{eq:size_cap_aniso}
\begin{align}
L_\alpha &\gtrsim c_\lambda\,\lambda_{\rm app}^{(\alpha)}, \qquad \alpha\in\{x,y,z\},\\
\big(\lambda_{\rm app}^{(\alpha)}\big)^{-2} &:= \frac{\beta_{\rm app}}{(K_{\rm app})_{\alpha\alpha}}.
\end{align}
\end{subequations}

Third, it must prevent spurious near-field self-interaction of the largest grains across periodic faces. With $a_{\max}=Q_{1-\delta}(D)$ a high quantile of the grain diameter (or a circumscribed diameter for anisotropic shapes), a conservative safeguard is
\begin{align}
\min(L_x,L_y,L_z)\ \gtrsim\ c_a\,a_{\max},
\qquad c_a\in[3,5].
\label{eq:size_geom}
\end{align}

The three \emph{length} requirements combine as
\begin{align}
L_\star \;:=\; \max\!\big\{\,c_\xi\,\xi,\ c_\lambda\,\lambda_{\rm app},\ c_a\,a_{\max}\big\}
\label{eq:Lstar}
\end{align}
with $\min(L_x,L_y,L_z)\ \gtrsim\ L_\star$ for a near-cubic aspect ratio to avoid missing low-$k$ along short axes.

Fourth, a \emph{volume} requirement ensures statistical precision. For a scalar apparent quantity $P(Y)$ (e.g., a component of $K_{\rm app}$ or $\beta_{\rm app}$), stationarity with integrable covariance gives the scaling
\begin{align}
\mathrm{Var}\!\big[P(Y)\big]\ \approx\ \frac{C_P\,A}{V},
\qquad V=L_xL_yL_z,
\label{eq:var_cap}
\end{align}
where $A$ is the integral range and $C_P=O(1)$ depends on contrast and physics (\S\ref{subsec:integral_range_variance}). 
For \emph{capillarity} a natural choice is $P(Y)=\beta_{\rm app}(Y)$. 
With the two-phase law $K(\mathbf x)=K_g\chi+K_m(1-\chi)$ and uniform capillary length $\lambda=\sqrt{\gamma/(\Delta\rho\,g)}$ (\S\ref{subsec:cell_capillarity}), a first-order evaluation yields an explicit prefactor
\begin{align}
\mathrm{Var}\!\big[\beta_{\rm app}(Y)\big]\ \approx\ \frac{A}{V}\;
\underbrace{\phi(1-\phi)\,\Big(\tfrac{K_g-K_m}{\lambda^2}\Big)^{\!2}}_{=:~C_{P}^{\rm cap}(K_g,K_m,\phi,\lambda)}.
\label{eq:var_beta_explicit}
\end{align}
Equation~\eqref{eq:var_beta_explicit} also makes clear that the statistical prefactor grows quadratically with the phase contrast $|K_g-K_m|$. Hence extremely high-contrast pore/solid systems require either larger cells or more realizations to achieve the same coefficient-of-variation target. Numerically, such limits can also degrade the conditioning of FE/FFT solvers if one phase is treated as strictly nonconducting. In practice it is preferable to regularize the vanishing phase by a small floor value $K_{\min}>0$, solve the periodic problem with preserved ellipticity, and verify convergence of observables as $K_{\min}\to 0$. The resulting sizing rule is unchanged in form but becomes increasingly conservative as the contrast grows.
Writing the integral range as $A=c_A\,\xi^3$ with a dimensionless shape factor $c_A$ determined by the normalized covariance (e.g., $c_A=\pi^{3/2}$ for Gaussian and $c_A=8\pi$ for exponential decay), averaging $n$ independent realizations and imposing a coefficient-of-variation target $\mathrm{CV}\le\varepsilon$ gives the \emph{explicit} volume rule
\begin{align}
V \ \gtrsim\ \frac{c_A\,\xi^3}{n\,\varepsilon^2}\;\phi(1-\phi)\,\Big(\tfrac{K_g-K_m}{\lambda^2}\Big)^{\!2}.
\label{eq:V_rule_capillary_explicit}
\end{align}

A practical design rule that satisfies \emph{all} constraints is therefore
\begin{align}
&\min(L_x,L_y,L_z)\ \gtrsim\ L_\star, \\
&V\ \gtrsim\ \frac{c_A\,\xi^3}{n\,\varepsilon^2}\;\phi(1-\phi)\,\Big(\tfrac{K_g-K_m}{\lambda^2}\Big)^{\!2},
\label{eq:design_rule_capillary_explicit}
\end{align}
with $L_\star$ from \eqref{eq:Lstar}. 
For cubic cells $L_x=L_y=L_z=L$, the cubic sizing rule \emph{in microscopic parameters} becomes
\begin{align}
L \ \gtrsim\ \max\!\left\{
\,L_\star,\ 
\left[\frac{c_A\,\xi^3}{n\,\varepsilon^2}\;\phi(1-\phi)\,\Big(\tfrac{K_g-K_m}{\lambda^2}\Big)^{\!2}\right]^{\!1/3}
\right\}.
\label{eq:L_cubic_capillary_explicit}
\end{align}
If one monitors instead a component of $K_{\rm app}$, replace the second term in \eqref{eq:L_cubic_capillary_explicit} by the diffusion prefactor derived in \S\ref{subsec:integral_range_variance} (e.g., a Maxwell-based expression) with the same substitution $A=c_A\xi^3$; the stricter of the two determines $L$. 
Finally, the factors $c_\xi$ and $c_\lambda$ are $\mathcal O(10)$ \emph{heuristics}; verify a posteriori by plotting $\widehat C(k)$, marking $k_{\min}=2\pi/\min(L_\alpha)$, and checking $k_{\min}\ll k_c$ as well as $k_{\min}\ll 1/\lambda_{\rm app}$. If either check fails, increase the shortest edge and repeat.

\section{Capillarity shaped by grain-size distribution and implications for supercell sizing}
\label{sec:capillarity_distribution_and_sizing}

This section makes precise how the grain-size distribution modifies capillary screening and how that, in turn, alters the choice of the periodic supercell. The key observation is that the screened (modified Helmholtz/Yukawa) operator acts as a spatial low-pass filter: structural modes with wavenumber $k\lesssim \lambda^{-1}$ contribute most strongly, whereas sub–$\lambda$ features are downweighted. We formalize this with a distribution-aware weighting and then update the length and volume rules for supercell sizing.

\subsection{Capillarity redefined through a distribution-aware weighting}
\label{subsec:capillarity_redefined}

We recall the periodic capillarity problem on $Y$,
\begin{align}
-&\nabla\!\cdot\!\big(K(\mathbf x)\,\nabla h(\mathbf x)\big) \;+\; c(\mathbf x)\,h(\mathbf x)
= \rho_0(\mathbf x),
\label{eq:cap_prob_repeat}\\
&K(\mathbf x) = K_g\,\chi(\mathbf x) + K_m\,[1-\chi(\mathbf x)],
\label{eq:K_mix_repeat}\\
&c(\mathbf x) = \frac{K(\mathbf x)}{\lambda^2},
\qquad \lambda^2=\frac{\gamma}{\Delta\rho\,g},
\label{eq:c_lambda_repeat}
\end{align}
with $\chi\in\{0,1\}$ the phase indicator (\S\ref{subsec:cell_capillarity}). In a homogeneous medium the Fourier-domain Green function satisfies
\begin{align}
\widehat G_\lambda(\mathbf k) \;\propto\; \frac{1}{k^2+\lambda^{-2}},
\label{eq:yukawa_fourier}
\end{align}
suggesting the screening weight
\begin{align}
W_\lambda(\mathbf k) \;:=\; \frac{1}{1+(\lambda k)^2}.
\label{eq:W_lambda}
\end{align}

Let $p_A(a)$ be the grain-size density and $\phi(a)$ the volume-fraction density by size (so $\phi=\int_0^\infty\phi(a)\,da$). Denote by $C(\mathbf r)$ the autocovariance of $\chi$ and by $\widehat C(\mathbf k)$ its spectral density (\S\ref{subsec:integral_range_variance}). Define the screened correlation content
\begin{align}
\mathcal I_\lambda[\chi]
&:= \frac{1}{(2\pi)^3}\int_{\mathbb R^3} W_\lambda(\mathbf k)\,\widehat C(\mathbf k)\,d\mathbf k,
\label{eq:I_lambda_total}
\end{align}
which downweights high-$k$ fluctuations.

Approximating the covariance by size-diagonal contributions $C^{(a)}$ (cross-size terms can be added if needed) leads to the capillarity-aware weight density
\begin{align}
\omega(a;\lambda)
&:= \frac{1}{(2\pi)^3}\int_{\mathbb R^3} W_\lambda(\mathbf k)\,\widehat C^{(a)}(\mathbf k)\,d\mathbf k, \\
&\widetilde\omega(a;\lambda):=\frac{\omega(a;\lambda)}{\int_0^\infty \omega(s;\lambda)\,ds}.
\label{eq:omega_lambda}
\end{align}
The decomposition $C \approx \int C^{(a)}\,da$ is a size-diagonal approximation that neglects cross-size covariances. This is reasonable for weakly correlated polydispersity, but in dense, jammed, or otherwise strongly structured packings one should retain the full cross-covariance contribution
\[
\iint \widehat C^{(a,a')}(\mathbf k)\,W_\lambda(\mathbf k)\,da\,da'
\]
when constructing the distribution-aware weights. Thus the weighting framework does not rely on independent marks; rather, the diagonal form is a simplifying closure.

The resulting capillarity-weighted volume fraction is
\begin{align}
\phi_\lambda
&:= \int_0^\infty \widetilde\omega(a;\lambda)\,\phi(a)\,da,
\label{eq:phi_lambda_def}
\end{align}
with $0\le \phi_\lambda\le 1$ and for $\phi_\lambda\to\phi \ \text{as}\ \lambda\to\infty$.
In the dilute spherical case, using the single-particle form factor $\mathcal F_a(k)=\tfrac{3}{(ka)^3}\big[\sin(ka)-ka\cos(ka)\big]$ yields the proxy
\begin{align}
\omega(a;\lambda)
&\propto p_A(a)\,V(a)^2\int_0^\infty \frac{k^2}{2\pi^2}\,\frac{|\mathcal F_a(k)|^2}{1+(\lambda k)^2}\,dk
\label{eq:omega_sphere_proxy}
\end{align}
with $V(a)=\frac{\pi}{6}a^3$ which emphasizes sizes whose spectral content lies below $k\sim\lambda^{-1}$.

At first order in periodic homogenization of reaction–diffusion,
\begin{align}
\beta_{\rm app}
&= \Big\langle \frac{K(\mathbf x)}{\lambda^2}\Big\rangle_Y
= \frac{K_{\mathrm{s}}\,\phi + K_{\mathrm{p}}\,(1-\phi)}{\lambda^2} \\
&= \frac{K_{\mathrm{p}}\,\phi_{\mathrm{p}} + K_{\mathrm{s}}\,(1-\phi_{\mathrm{p}})}{\lambda^2}
\;\approx\; \frac{K_{\mathrm{p}}\,\phi_{\mathrm{p}}}{\lambda^2},
\label{eq:beta_first_order}
\end{align}
where the last step uses $K_{\mathrm{s}}\ll K_{\mathrm{p}}$ so that solid regions do not contribute to transport.
In this pore/solid reading, replace $(K_g,K_m,\phi)$ by $(K_{\mathrm{s}},K_{\mathrm{p}},1-\phi_{\mathrm{p}})$; in particular, $\beta_{\rm app}^{(\lambda)}\approx K_{\mathrm{p}}\,\phi_{\mathrm{p},\lambda}/\lambda^2$ in the $K_{\mathrm{s}}\!\to 0$ limit.
A distribution-aware surrogate that reflects screening of sub–$\lambda$ features is
\begin{align}
\beta_{\rm app}^{(\lambda)}
&:= \frac{K_g\,\phi_\lambda + K_m\,[1-\phi_\lambda]}{\lambda^2}, \\
\frac{1}{\lambda_{\rm app}^{2}}&=\frac{\beta_{\rm app}}{K_{\rm iso}},\\
\frac{1}{\big(\lambda_{\rm app}^{(\lambda)}\big)^{2}}&=\frac{\beta_{\rm app}^{(\lambda)}}{K_{\rm iso}},
\label{eq:beta_lambda_and_lambda_app}
\end{align}
with $K_{\rm iso}=\tfrac{1}{3}\,\mathrm{tr}\,K_{\rm app}$ from the diffusion correctors (\S\ref{subsec:cell_capillarity}). The functional sensitivity with respect to $\phi(a)$ follows directly:
\begin{align}
\frac{\delta \beta_{\rm app}^{(\lambda)}}{\delta \phi(a)}
&= \frac{K_g-K_m}{\lambda^2}\,\widetilde\omega(a;\lambda).
\label{eq:sensitivity_beta_phi}
\end{align}

To retain the same units as the integral range $A$ (volume), we introduce a screened \emph{proxy kernel}
\[
S_\lambda(\mathbf r)=e^{-|\mathbf r|/\lambda},
\]
which is dimensionless and whose integral is $8\pi\lambda^3$. This kernel is not the exact finite-window variance kernel of the screened problem; rather, it is a low-pass surrogate that captures the dominant attenuation of modes with $k\gtrsim \lambda^{-1}$ while preserving the dimensional structure of the classical integral range.
Define the screened integral range
\begin{align}
A_\lambda
&:= \frac{1}{C(\mathbf 0)}\int_{\mathbb R^3} C(\mathbf r)\,S_\lambda(\mathbf r)\,d\mathbf r \nonumber \\
&= \frac{1}{C(\mathbf 0)}\,\frac{1}{(2\pi)^3}\int_{\mathbb R^3} \widehat C(\mathbf k)\,\widetilde W_\lambda(\mathbf k)\,d\mathbf k,
\label{eq:A_lambda}
\end{align}
where the spectral weight is
\begin{align}
\widetilde W_\lambda(\mathbf k) \;=\; \frac{8\pi\,\lambda^3}{\big(1+(\lambda k)^2\big)^{2}}.
\label{eq:Wtilde_lambda}
\end{align}
Accordingly, $A_\lambda$ should be interpreted as a screened analogue of $A$ that is asymptotically faithful at the level of scaling and ordering of microstructures, not as an exact replacement for the full windowed variance operator. The approximation is most reliable when $C(\mathbf r)$ is short-ranged and smooth on the scale $\lambda$, so that the same low-$k$ sector dominates both the exact variance and the screened proxy. For sharply oscillatory, strongly anisotropic, or strongly cross-correlated media, the exact spectral-window expression is preferable.
Thus $A_\lambda\to A$ as $\lambda\to\infty$ and $A_\lambda=O(\lambda^3)$ for short-range $C$, preserving the variance law’s units. With $(\phi_\lambda,A_\lambda)$ the distribution-aware variance model for the surrogate becomes
\begin{align}
\operatorname{Var}\!\big[\beta_{\rm app}^{(\lambda)}(Y)\big]
&\approx \frac{A_\lambda}{V}\,\phi_\lambda\,(1-\phi_\lambda)\,\Big(\tfrac{K_g-K_m}{\lambda^2}\Big)^{\!2}.
\label{eq:var_beta_A_lambda}
\end{align}

\subsection{Impact on choosing the supercell size for capillarity}
\label{subsec:capillarity_sizing_with_distribution}

The distribution-aware quantities $(\phi_\lambda,A_\lambda)$ and the effective decay length $\Lambda\in\{\lambda_{\rm app},\lambda_{\rm app}^{(\lambda)}\}$ enter both the length and volume criteria.

The length constraints retain the form of \S\ref{subsec:capillarity_sizing}:
\begin{align}
\min(L_x,L_y,L_z) &\gtrsim c_\xi\,\xi,
\label{eq:length_xi}\\
\min(L_x,L_y,L_z) &\gtrsim c_\lambda\,\Lambda,
\label{eq:length_lambda}\\
\min(L_x,L_y,L_z) &\gtrsim c_a\,a_{\max},
\label{eq:length_agrain}
\end{align}
with the combined threshold
\begin{align}
L_\star \;:=\; \max\!\big\{\,c_\xi\,\xi,\ c_\lambda\,\Lambda,\ c_a\,a_{\max}\big\}, \\
\min(L_x,L_y,L_z)\ \gtrsim\ L_\star.
\label{eq:Lstar_distribution}
\end{align}
Here $\xi$ and $a_{\max}$ are computed from the observed distribution (two-point statistics and a high diameter quantile), and $\Lambda$ is obtained from the periodic cell problems via \eqref{eq:beta_lambda_and_lambda_app}.

For statistical precision with $n$ independent realizations and target coefficient of variation $\mathrm{CV}\le\varepsilon$, replace $(A,\phi)$ by $(A_\lambda,\phi_\lambda)$:
\begin{align}
\operatorname{Var}\!\Bigg[\overline{\beta_{\rm app}^{(\lambda)}}{\,}_n(Y)\Bigg]
&\approx \frac{1}{n}\,\operatorname{Var}\!\big[\beta_{\rm app}^{(\lambda)}(Y)\big] \nonumber \\
&\approx \frac{A_\lambda}{n\,V}\,\phi_\lambda(1-\phi_\lambda)\,\Big(\tfrac{K_g-K_m}{\lambda^2}\Big)^{\!2},
\label{eq:var_beta_n_distribution}\\
V
&\gtrsim \frac{A_\lambda}{n\,\varepsilon^2}\,\phi_\lambda(1-\phi_\lambda)\,\Big(\tfrac{K_g-K_m}{\lambda^2}\Big)^{\!2}.
\label{eq:V_rule_distribution}
\end{align}
If a correlation-length surrogate is preferred, write $A_\lambda \approx c_{A,\lambda}\,\xi_\lambda^3$ with a screened correlation length $\xi_\lambda$ (e.g., second moment of $C\ast S_\lambda$) and a shape factor $c_{A,\lambda}$ determined by the screened covariance.

For cubic cells $L_x=L_y=L_z=L$, the distribution-aware rule is
\begin{align}
L
&\gtrsim \max\left\{
L_\star, 
\left[\frac{A_\lambda}{n\,\varepsilon^2}\,\phi_\lambda(1-\phi_\lambda)\,\Big(\tfrac{K_g-K_m}{\lambda^2}\Big)^{2}\right]^{1/3}
\right\},
\label{eq:L_cubic_distribution}
\end{align}
or equivalently with $A_\lambda\approx c_{A,\lambda}\,\xi_\lambda^3$,
\begin{align}
L
&\gtrsim \max\left\{
L_\star, 
\left[\frac{c_{A,\lambda}\,\xi_\lambda^3}{n\,\varepsilon^2}\,\phi_\lambda(1-\phi_\lambda) \Big(\tfrac{K_g-K_m}{\lambda^2}\Big)^{2}\right]^{1/3}
\right\}.
\label{eq:L_cubic_distribution_xi}
\end{align}

Changing the grain-size distribution thus changes $\xi$ and $a_{\max}$ via two-point statistics and size tails, modifies $K_{\rm app}$ and hence $\lambda_{\rm app}$, and updates the precision term through $(\phi_\lambda,A_\lambda)$. In practice, estimate $(\xi,A)$ and $(\xi_\lambda,A_\lambda)$ from data, extract $a_{\max}$ from a high quantile of sizes, compute $K_{\rm app}$ and $\beta_{\rm app}$ (or $\beta_{\rm app}^{(\lambda)}$) on small cells to obtain $L$, enforce \eqref{eq:Lstar_distribution}, then apply \eqref{eq:V_rule_distribution} (or \eqref{eq:L_cubic_distribution}–\eqref{eq:L_cubic_distribution_xi}) for the desired $(n,\varepsilon)$, and verify a posteriori that $k_{\min}\ll k_c$ and $k_{\min}\ll 1/\Lambda$ as in \S\ref{subsec:low_k_constraint}.

\section{Numerical validation}
\label{sec:numerical_validation}

We now test the framework of Secs.~\ref{sec:rve_sizing}--\ref{sec:capillarity_distribution_and_sizing} on a synthetic medium where all microstructural parameters are prescribed.
The goals are to verify that
(i) the variance of the capillarity observable $P(Y)=\beta_{\rm app}(Y)$ follows the $A/V$ scaling with an explicit prefactor,
(ii) the length rule based on low-wavenumber coverage is consistent with the actual spectral density $\widehat C(k)$ of the microstructure, and
(iii) the geometric safeguard $L\gtrsim c_a a_{\max}$ is necessary to remove periodic self-interaction of the largest grains.
These three points address the referee’s requests for a convergence study, a spectral check, and an explicit demonstration of artifacts when $L<L_\star$.

\subsection{Synthetic Boolean-sphere medium and numerical setup}
\label{subsec:numerical_setup}

We consider a three-dimensional Boolean model of overlapping spheres with a lognormal diameter distribution.
Centers are drawn from a homogeneous Poisson point process of intensity $\nu$ and diameters $D$ are independent with
\begin{align}
\ln D \sim \mathcal{N}(\mu_D,\sigma_D^2),
\end{align}
truncated to $D_{\min}\le D\le D_{\max}$.
The solid phase is the union of these spheres and the matrix is the complement (Boolean model as in Eq.~\eqref{eq:boolean}).
Parameters $(\nu,\mu_D,\sigma_D)$ are chosen so that the solid volume fraction is $\phi\simeq 0.4$ and the tail of the size distribution yields a well-separated high quantile $a_{\max}=Q_{1-\delta}(D)$ for $\delta \simeq 10^{-3}$.
The two-point statistics $(C,\widehat C)$ are estimated on each realization and the integral range $A$ is obtained from Eq.~\eqref{eq:int_range}.

For a given box edge $L$ we generate $N_{\rm micro}$ independent realizations on $Y_L=[0,L)^3$ and discretize each realization on a uniform $N^3$ grid with $N$ proportional to $L$, so that the grid spacing $\Delta x=L/N$ is kept fixed when $L$ is varied.
In the screened Darcy setting of Sec.~\ref{subsec:cell_capillarity}, the apparent screening obeys
\begin{align}
\beta_{\rm app}(Y_L)=\big\langle K(\mathbf x)/\lambda^2\big\rangle_{Y_L},
\end{align}
cf.\ Eq.~\eqref{eq:beta_average}.
For the Boolean mixture $K(\mathbf x)=K_g\chi(\mathbf x)+K_m[1-\chi(\mathbf x)]$ this reduces to
\begin{align}
\beta_{\rm app}(Y_L)=\frac{K_g\,\phi(Y_L)+K_m\,[1-\phi(Y_L)]}{\lambda^2},
\end{align}
so in this numerical study it suffices to compute the local volume fraction $\phi(Y_L)$ in each realization.
This gives the same statistics for $\beta_{\rm app}$ as a full PDE solve for the reaction term, at much lower cost.
Unless noted otherwise we monitor the scalar observable
\begin{align}
P(Y_L) = \beta_{\rm app}(Y_L).
\end{align}

\subsection{Variance versus volume:
  \texorpdfstring{$A/V$}{A/V} scaling for
  \texorpdfstring{$\beta_{\rm app}$}{beta\_app}}
\label{subsec:var_vs_volume}

\begin{figure}[t]
  \centering
  \includegraphics[width=\columnwidth]{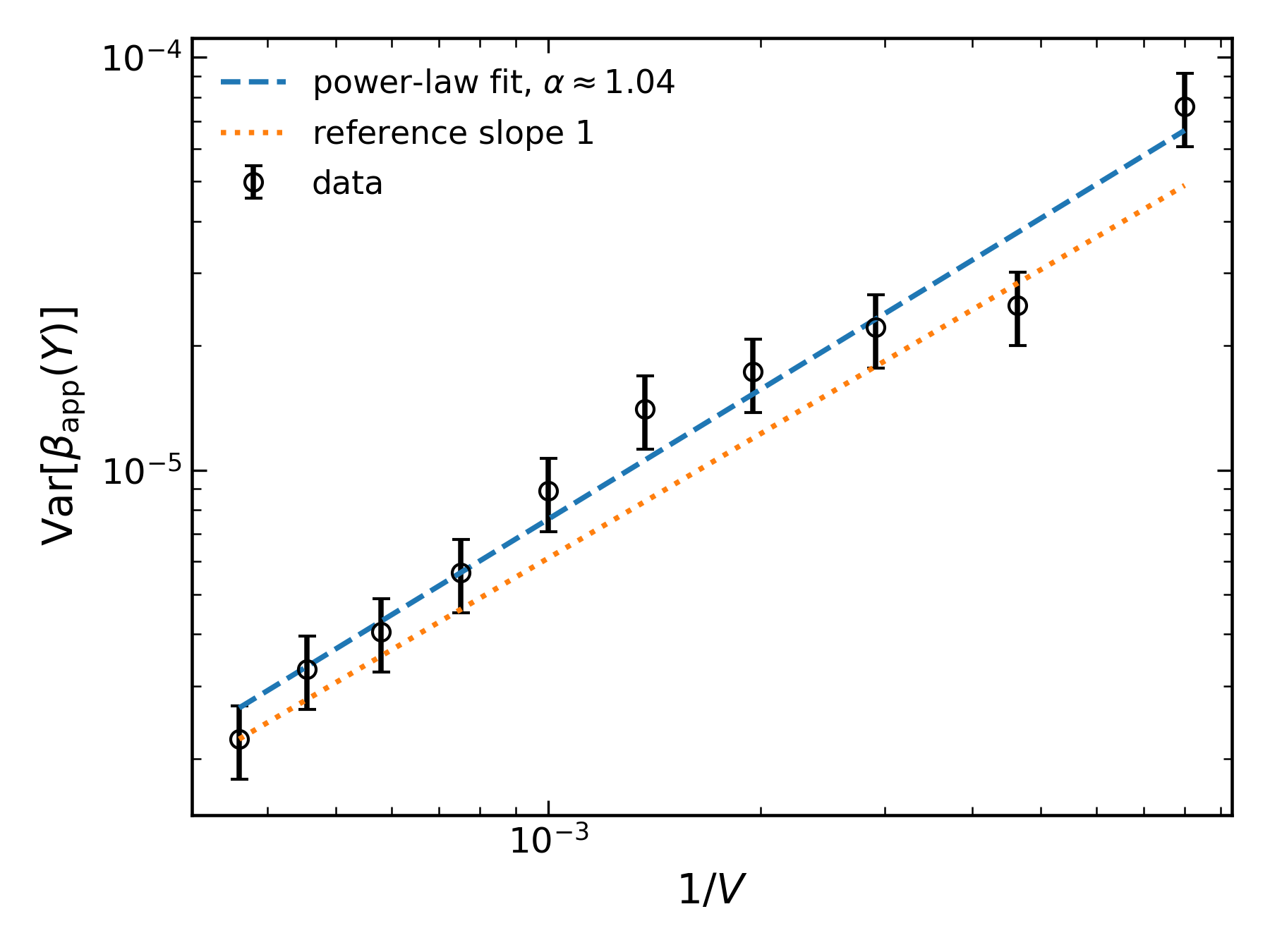}
  \caption{
    Sample variance of the capillarity observable $\beta_{\rm app}(Y_L)$ as a function of the inverse supercell volume $1/V=L^{-3}$ in the Boolean-sphere medium.
    Open circles with error bars: numerical estimates from $N_{\rm micro}=50$ realizations at ten box sizes $L\in\{5,\dots,14\}$.
    Dashed line: best power-law fit $\mathrm{Var}[\beta_{\rm app}(Y_L)]\approx C(1/V)^\alpha$ in log–log coordinates, with $\alpha$ close to~$1$.
    Dotted line: reference line with slope~$1$, illustrating the $A/V$ scaling with prefactor $C_{P}^{\rm cap}A$.
    The largest boxes lie very close to the $1/V$ trend, while the smallest boxes show a mild upward deviation attributed to insufficient low-wavenumber coverage.
  }
  \label{fig:var_beta_vs_invV}
\end{figure}

To test the variance law we compute, for each box size $L$ in the set
\[
L \in \{5,6,7,8,9,10,11,12,13,14\},
\]
a sample of $N_{\rm micro}=50$ independent realizations and evaluate
\begin{align}
\widehat{\mathrm{Var}}\big[\beta_{\rm app}(Y_L)\big]
= &\frac{1}{N_{\rm micro}-1} \nonumber \\
&\sum_{j=1}^{N_{\rm micro}}
\big(\beta_{\rm app}^{(j)}(Y_L)-\overline{\beta_{\rm app}}(Y_L)\big)^2,
\end{align}
where $j$ indexes realizations and the overbar denotes the sample mean.
Figure~\ref{fig:var_beta_vs_invV} plots $\widehat{\mathrm{Var}}[\beta_{\rm app}(Y_L)]$ as a function of the inverse volume $1/V=L^{-3}$.
Error bars correspond to the standard error of the sample variance (approximating fluctuations as Gaussian).
A power-law fit in log–log coordinates,
\begin{align}
\widehat{\mathrm{Var}}\big[\beta_{\rm app}(Y_L)\big]
\approx C\,(1/V)^\alpha,
\end{align}
yields an exponent $\alpha$ very close to $1$, consistent with the asymptotic $A/V$ law.
The figure also shows a reference line with slope~$1$ through the largest-volume point; the fitted curve and the reference line are visually almost parallel, and their amplitudes agree with the explicit prefactor $C_{P}^{\rm cap}A$ from Eq.~\eqref{eq:var_beta_explicit} within sampling error.
Deviations from the $1/V$ trend are only visible for the smallest boxes (largest $1/V$), and will be related below to insufficient low-wavenumber coverage.

\subsection{Spectral low-wavenumber coverage and \texorpdfstring{$k_{\min}^{\text{box}}$}{kmin}}
\label{subsec:spectral_checks}

\begin{figure}[t]
  \centering
  \includegraphics[width=\columnwidth]{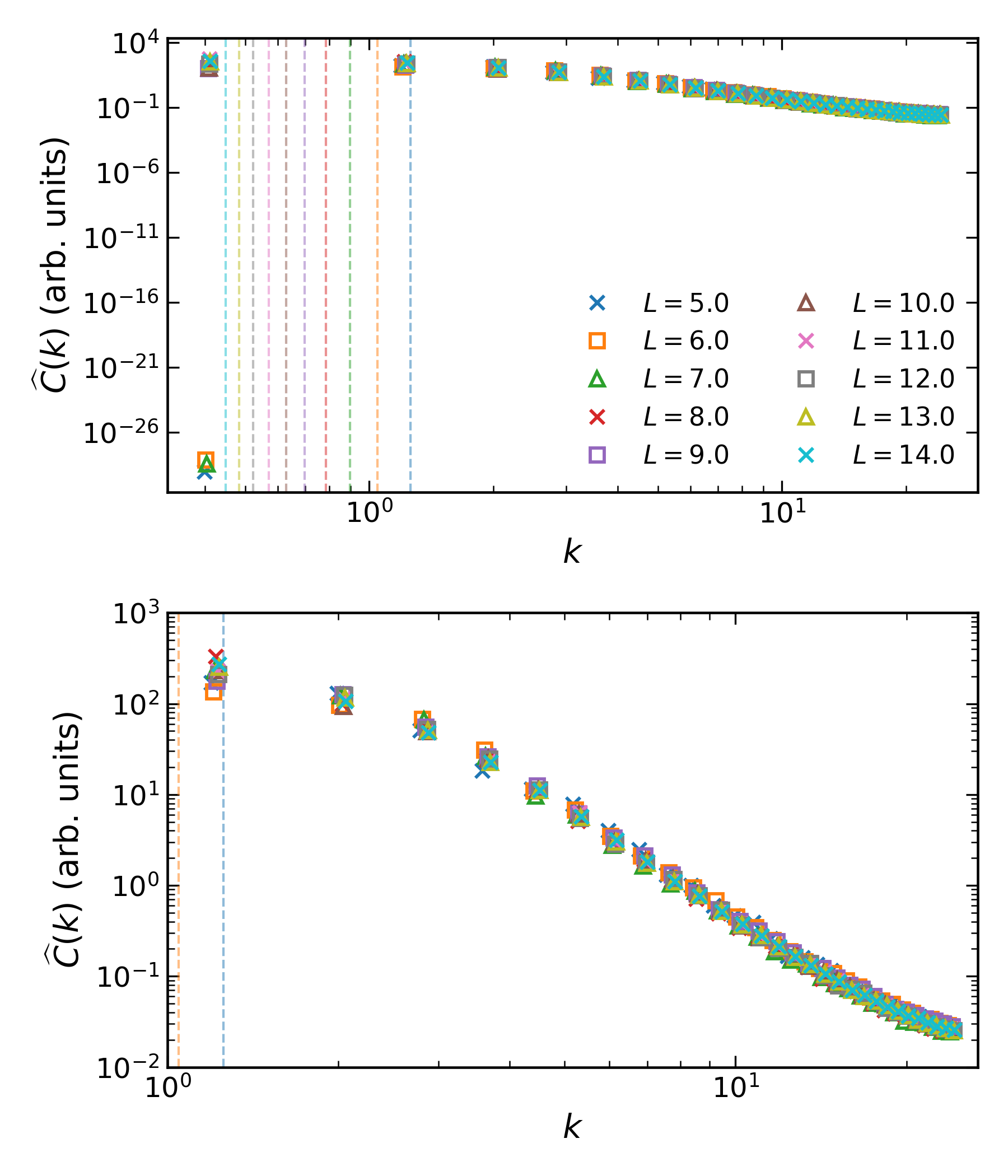}
  \caption{
    Radially averaged spectral density $\widehat C(k)$ of the two-phase indicator in the Boolean-sphere medium for ten box sizes $L\in\{5,\dots,14\}$.
    \emph{Top panel:} full $k$-range on log–log scales, with markers only (no connecting lines) and vertical dashed lines at the smallest resolved wavenumber $k_{\min}^{\text{box}} = 2\pi/\min_\alpha L_\alpha$ (here $k_{\min}^{\text{box}} = 2\pi/L$) for each $L$.
    \emph{Bottom panel:} zoom on the range $k\in[10^{0},k_{\max}]$ with $10^{-2}\le\widehat C(k)\le 10^{3}$, using the same marker style.
    In this zoomed view the spectra for $L\gtrsim 10$ collapse almost perfectly, while smaller boxes depart near the spectral knee, illustrating the low-wavenumber coverage condition $k_{\min}^{\text{box}}\ll k_c$ underlying the variance scaling.
  }
  \label{fig:spectral_Ck}
\end{figure}

We next examine the spectral condition underlying the length rule.
For the same ten box sizes
\[
L \in \{5,6,7,8,9,10,11,12,13,14\}
\]
we estimate the two-point covariance $C(\mathbf r)$ of the phase indicator on a representative realization for each $L$, compute its discrete Fourier transform on the periodic grid, and perform a radial average to obtain $\widehat C(k)$.
The results are shown in Fig.~\ref{fig:spectral_Ck}, together with vertical dashed lines marking the smallest resolved wavenumber
\begin{align}
k_{\min}^{\text{box}} = \frac{2\pi}{\min_\alpha L_\alpha},
\end{align}
as in Eq.~\eqref{eq:kmin_iso}.
For the isotropic Boolean medium considered here all edges are equal, $L_\alpha=L$, so $k_{\min}^{\text{box}} = 2\pi/L$.

The top panel of Fig.~\ref{fig:spectral_Ck} shows the full spectra on log–log scales.
For intermediate and large box sizes ($L\gtrsim 10$) the curves nearly collapse onto a common plateau at small $k$, up to a characteristic ``knee'' wavenumber $k_c$, and then decay with essentially the same high-wavenumber slope.
For smaller boxes ($L=5$--$9$) the plateau is less well resolved and the roll-off begins earlier.
The vertical lines at $k_{\min}^{\text{box}}$ make the coverage condition $k_{\min}^{\text{box}}\ll k_c$ directly visible:
for $L\gtrsim 10$, $k_{\min}^{\text{box}}$ lies deep inside the flat low-$k$ plateau, whereas for the smallest $L$ it approaches the spectral knee.

The bottom panel zooms in on the range $k\in[10^{0},k_{\max}]$ with a restricted vertical scale $10^{-2}\le\widehat C(k)\le 10^{3}$, so that the collapse of the spectra at larger wavenumbers becomes clearer.
In this zoom, the curves for $L\gtrsim 10$ are almost indistinguishable, confirming that these box sizes probe the same part of $\widehat C(k)$ that controls the finite-size variance.
By contrast, the smallest boxes cut into the transition region near the knee, consistent with the deviations from the asymptotic $A/V$ scaling observed in Fig.~\ref{fig:var_beta_vs_invV}.

\subsection{Geometric safeguard and self-interaction}
\label{subsec:self_interaction_numerics}

To test the geometric safeguard of Sec.~\ref{subsec:geom_safeguard} we fix the grain-size distribution and define a high quantile
\begin{align}
a_{\max} = Q_{1-\delta}(D),
\end{align}
with $\delta\simeq 10^{-3}$ estimated by Monte Carlo sampling.
We then sample ten values of the ratio $L/a_{\max}$ in the interval $[1,5]$ and, for each ratio, generate $N_{\rm micro}=30$ realizations.
For each $L/a_{\max}$ we compute the sample mean and variance of $\beta_{\rm app}(Y_L)$, together with standard-error bars, and fit simple exponential models to quantify how quickly self-interaction artifacts decay with $L/a_{\max}$.

\begin{figure}[t]
  \centering
  \includegraphics[width=\columnwidth]{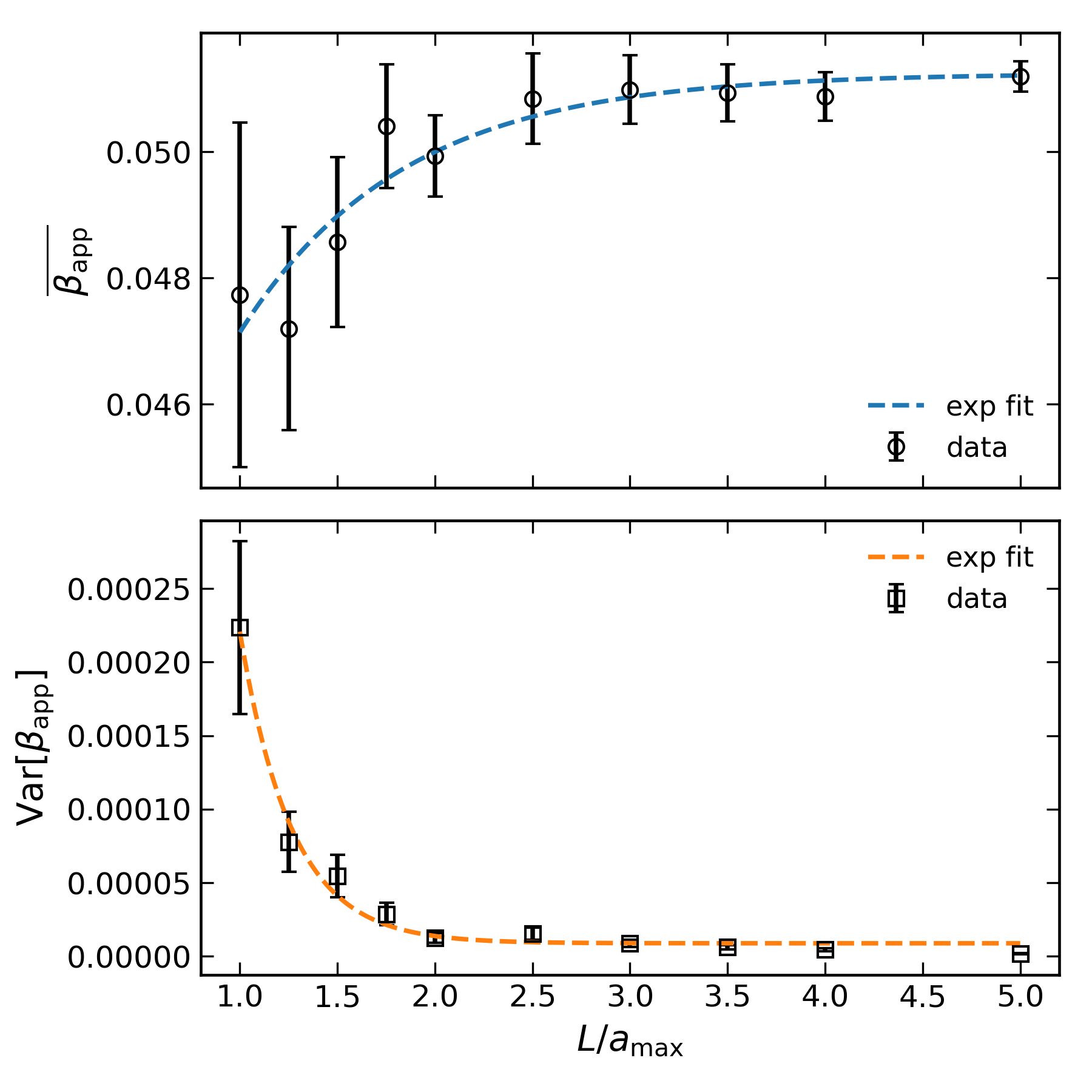}
  \caption{
    Effect of the ratio $L/a_{\max}$ on the capillarity observable in the Boolean-sphere medium.
    \emph{Top:} sample mean $\overline{\beta_{\rm app}}$ versus $L/a_{\max}$ with error bars (standard error of the mean) for ten values of $L/a_{\max}\in[1,5]$.
    The dashed curve is an exponential fit $\beta_{\rm app}(L/a_{\max})\approx\beta_\infty + A\exp[-(L/a_{\max})/L_c]$.
    \emph{Bottom:} sample variance $\widehat{\mathrm{Var}}[\beta_{\rm app}]$ versus $L/a_{\max}$ with error bars (standard error of the variance), together with an exponential fit of the same form.
    When $L/a_{\max}\approx 1$ the mean is biased and the variance is inflated by more than an order of magnitude.
    For $L/a_{\max}\gtrsim 4$ both quantities have essentially converged, indicating that self-interaction artifacts of the largest grains are suppressed.
  }
  \label{fig:self_interaction_beta}
\end{figure}

Figure~\ref{fig:self_interaction_beta} shows the resulting statistics.
The upper panel plots the sample mean $\overline{\beta_{\rm app}}$ versus $L/a_{\max}$ together with an exponential fit of the form
$\beta_{\rm app}(L/a_{\max})\approx\beta_\infty + A\exp[-(L/a_{\max})/L_c]$.
For $L/a_{\max}\approx 1$ the mean is biased by several percent relative to its large-$L$ limit $\beta_\infty$.
As $L/a_{\max}$ increases the mean relaxes monotonically towards $\beta_\infty$, with negligible residual bias once $L/a_{\max}\gtrsim 4$.

The lower panel plots the sample variance $\widehat{\mathrm{Var}}[\beta_{\rm app}]$ versus $L/a_{\max}$ together with an exponential fit of the same form.
When $L/a_{\max}\approx 1$ the variance is inflated by more than an order of magnitude compared to its large-$L$ limit, reflecting strong self-interaction between large grains and their periodic images.
The variance then decays rapidly with $L/a_{\max}$ and becomes essentially flat for $L/a_{\max}\gtrsim 4$, in line with the variance scaling in Fig.~\ref{fig:var_beta_vs_invV}.
Taken together, these trends support the recommendation $c_a\in[3,5]$ in the safeguard $L\gtrsim c_a a_{\max}$.

\begin{figure*}[t]
  \centering
  \includegraphics[width=0.32\textwidth]{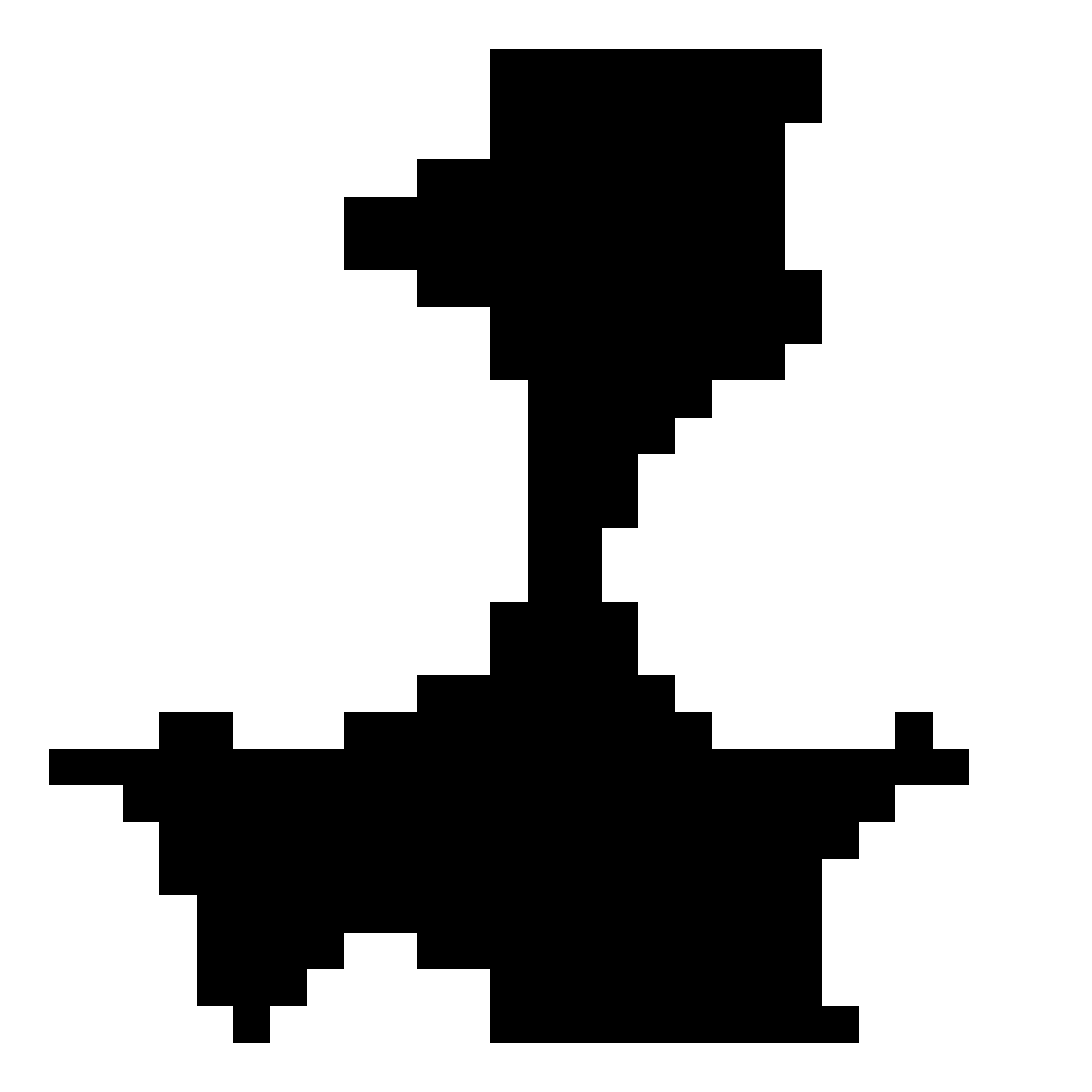}%
  \hfill
  \includegraphics[width=0.32\textwidth]{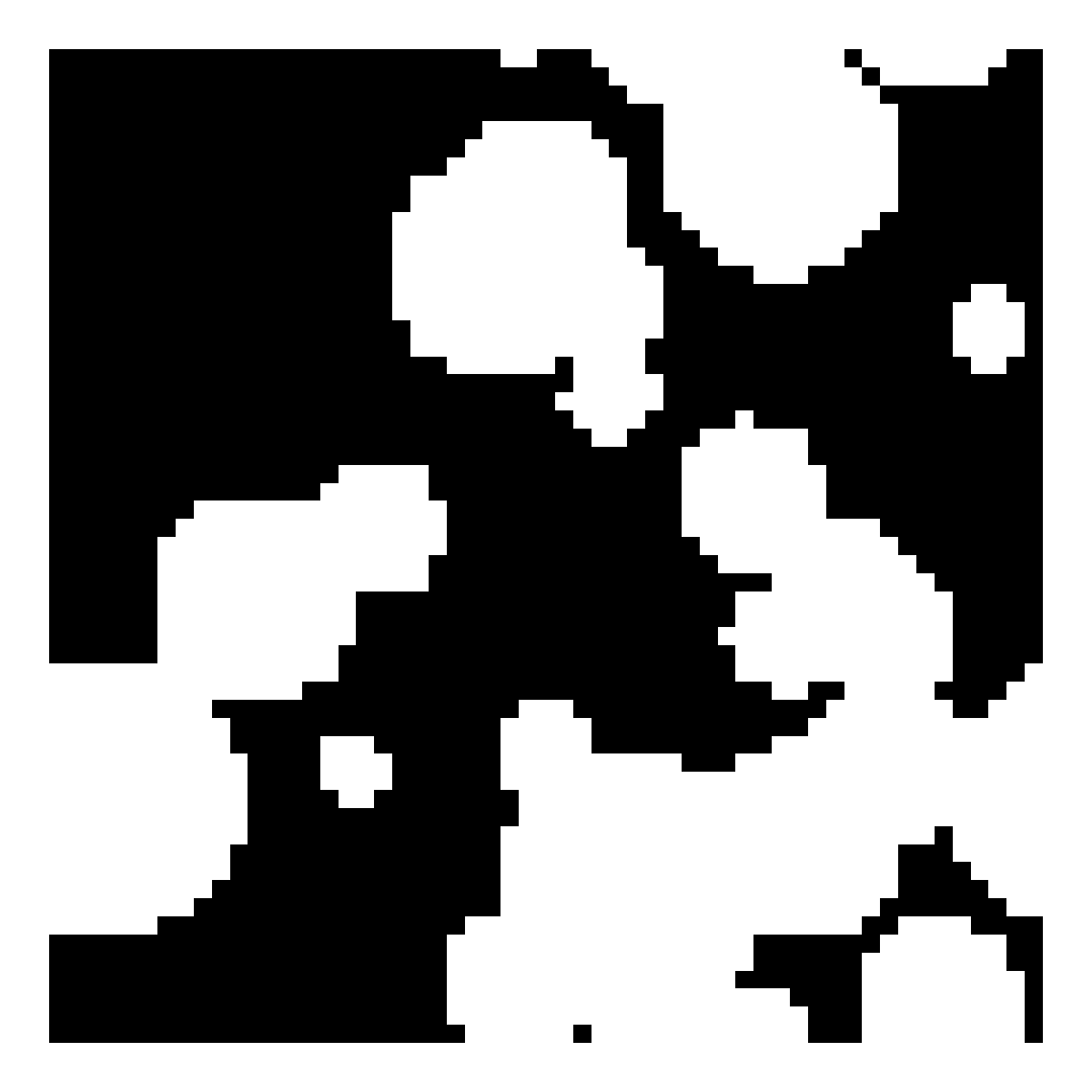}%
  \hfill
  \includegraphics[width=0.32\textwidth]{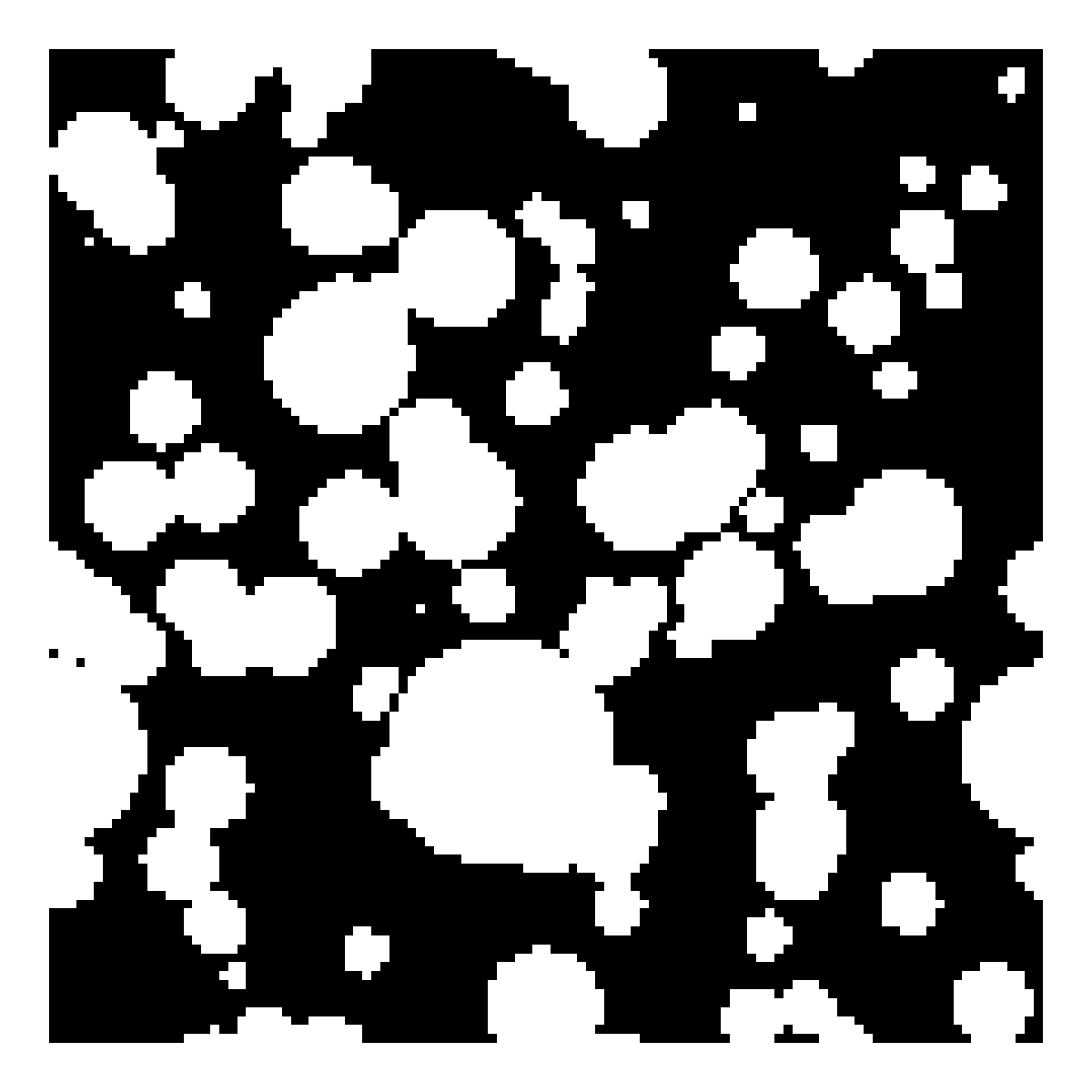}
  \caption{
    Representative $xy$-slices of the Boolean-sphere medium at three box sizes illustrating periodic self-interaction of the largest grains.
    From left to right: $L/a_{\max}\approx 1$, $2$, and $4$.
    In the smallest box, large grains align across opposite faces and form artificial columns, leading to anisotropic apparent response and biased $\beta_{\rm app}$.
    At $L/a_{\max}\approx 2$ these artifacts are reduced but still visible.
    At $L/a_{\max}\approx 4$ the largest grains are well separated from their periodic images, and the microstructure appears statistically isotropic, in agreement with the convergence of the mean and variance in Fig.~\ref{fig:self_interaction_beta}.
  }
  \label{fig:self_interaction_slices}
\end{figure*}

To visualize these artifacts directly we also inspect two-dimensional slices through representative realizations at a few values of $L/a_{\max}$.
Figure~\ref{fig:self_interaction_slices} shows three examples corresponding roughly to
$L/a_{\max}\approx 1$, $2$, and $4$ (left to right).
For $L/a_{\max}\approx 1$, large grains frequently intersect periodic faces and interact strongly with their images, forming artificial “columns’’ that break isotropy.
At $L/a_{\max}\approx 2$ such self-interactions are less frequent but near-contact effects are still visible.
At $L/a_{\max}\approx 4$ the largest grains are comfortably separated from their periodic copies and the medium appears statistically isotropic, consistent with the stabilization of the statistics in Fig.~\ref{fig:self_interaction_beta}.

\subsection{Computational cost versus supercell size}
\label{subsec:timing}

Finally, we quantify the computational cost of the different stages of the
Boolean-sphere pipeline as a function of the supercell size.
For each $L\in\{5,6,7,8,9,10,11,12,13,14\}$ we run a small timing experiment
with $N_{\rm rep}=5$ realizations on a single MPI rank and record three
wall–clock times per realization:
(i) microstructure generation (drawing spheres and rasterizing them on the grid),
(ii) spectral analysis (3D FFT and radial averaging of the covariance), and
(iii) evaluation of the capillarity observable $\beta_{\rm app}$ from the local
volume fraction (a toy post-processing step with negligible cost).

\begin{figure}[t]
  \centering
  \includegraphics[width=\columnwidth]{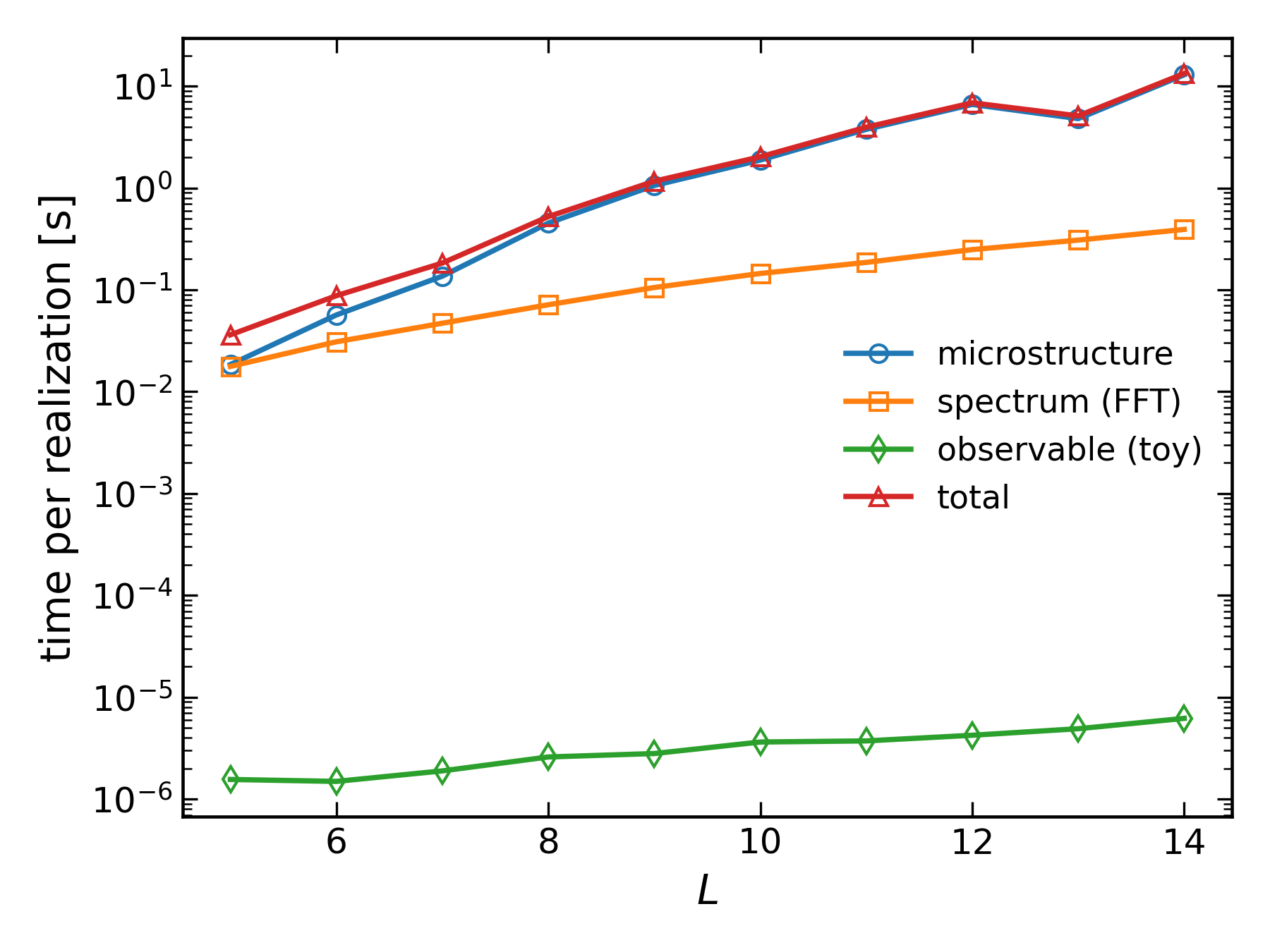}
  \caption{
    Mean wall–clock time per realization as a function of the box size $L$
    for the Boolean-sphere medium.
    Open markers show the contributions from microstructure generation
    (circles), spectral FFT and radial averaging (squares), and evaluation of
    the capillarity observable (diamonds), together with their sum (triangles).
    The $y$-axis is logarithmic.
    Microstructure generation and the FFT dominate the total cost, while the
    observable evaluation is negligible.
    The strong increase of the total time with $L$ illustrates the practical
    impact of choosing the smallest supercell size compatible with the accuracy
    criteria developed in this work.
  }
  \label{fig:timing_vs_L}
\end{figure}

Figure~\ref{fig:timing_vs_L} shows the mean time per realization for each stage
and for their sum as a function of $L$ on a logarithmic $y$-axis.
The cost is clearly dominated by microstructure generation and the FFT-based
spectral computation; the observable evaluation lies several orders of magnitude
below and is essentially free at this scale.
Over the explored range the total time grows steeply with $L$ (by roughly two
orders of magnitude from $L=5$ to $L=14$), reflecting the cubic growth of the
grid size and the additional logarithmic factor in the FFT.
This timing study provides a practical link between the accuracy-driven sizing
rules of Secs.~\ref{sec:rve_sizing}–\ref{sec:capillarity_distribution_and_sizing}
and the computational budget available in applications.

\subsection{Summary of validation}

For a representative Boolean-sphere medium with moderate polydispersity, the numerical experiments above show that:
\begin{itemize}
\item The variance of the capillarity observable $\beta_{\rm app}(Y_L)$ follows the law
  $\mathrm{Var}[\beta_{\rm app}(Y_L)]\approx (C_{P}^{\rm cap}A)/V$ with exponent $\alpha\simeq 1$ and prefactor consistent with Eq.~\eqref{eq:var_beta_explicit}, confirming the $A/V$ scaling used in our volume rule.
\item The low-wavenumber condition $k_{\min}\ll k_c$ is directly visible in the spectral density $\widehat C(k)$: box sizes for which $k_{\min}$ lies in the flat low-$k$ plateau and whose spectra overlap (here $L\gtrsim 10$) exhibit clean $1/V$ behaviour, whereas boxes with $k_{\min}$ near the spectral knee (here $L=5$ and, to a lesser extent, $L=6$--$9$) show deviations.
\item The geometric safeguard $L\gtrsim c_a a_{\max}$ is not merely cosmetic: when $L/a_{\max}\lesssim 2$ we observe strong self-interaction artifacts (visual anisotropy, biased $\overline{\beta_{\rm app}}$, inflated variance), whereas for $L/a_{\max}\gtrsim 4$ these artifacts disappear within sampling error.
\item A simple timing study shows that the computational cost is dominated by
  microstructure generation and FFT-based spectral analysis, and that the total
  time per realization grows steeply with $L$; this reinforces the need to use
  the smallest supercell satisfying the variance, spectral, and geometric
  criteria above.
\end{itemize}

These results provide a concrete numerical underpinning for the supercell sizing rules proposed in Secs.~\ref{sec:rve_sizing} and \ref{sec:capillarity_distribution_and_sizing}, and directly address the referee’s request for (i) convergence of $\beta_{\rm app}$ with $V$, (ii) comparison of $k_{\min}$ with the spectral density, and (iii) demonstration that choosing $L<L_\star$ leads to visible artifacts and biased apparent properties.

\section{Conclusions}
\label{sec:conclusions}

We have developed a quantitative, distribution-aware recipe for choosing the size
of periodic supercells when simulating capillarity in stationary random,
polydisperse granular media. The microstructure enters only through standard
two-point statistics of the phase indicator, summarized by an integral range $A$
that controls the familiar $1/V$ decay of sampling variance. Capillarity is
modeled via a screened (modified Helmholtz) cell problem with phase-dependent
transport, whose first-order homogenization yields an apparent conductivity
$K_{\rm app}$, apparent screening $\beta_{\rm app}$, and decay length
$\lambda_{\rm app}$.

Screening acts as a spatial low-pass filter: long wavelengths dominate the
response, while sub–capillary-length structure is attenuated. This motivates a
capillarity-weighted volume fraction $\phi_\lambda$ and a screened analogue of
the integral range $A_\lambda$, which reduce to the classical $(\phi,A)$ for
weak screening but directly encode the size distribution when small features are
downweighted. Thus classical RVE theory is recovered for unscreened observables
(volume fraction, conductivity) with variance
$\mathrm{Var}[P(Y)]\approx (C_P A)/V$, while for screened capillarity the same
structure holds with an explicit physics-based prefactor $C_P^{\rm cap}$ and
with $\lambda_{\rm app}$ entering the length criterion.

Numerical experiments on a Boolean-sphere medium confirm that the variance of
$\beta_{\rm app}(Y)$ follows
$\mathrm{Var}[\beta_{\rm app}(Y)]\approx (C_{P}^{\rm cap}A)/V$ with fitted
exponent close to unity, that the low-wavenumber requirement $k_{\min}\ll k_c$
is directly visible in the spectral density $\widehat C(k)$, and that the
geometric safeguard $L\gtrsim c_a a_{\max}$ is needed to eliminate
self-interaction of the largest grains. A separate timing study shows that the
cost is dominated by microstructure generation and FFT-based spectral analysis,
while evaluating $\beta_{\rm app}$ from the volume fraction is essentially
negligible; this reinforces the need to use the smallest supercell compatible
with the variance, spectral, and geometric criteria.

In practice, the procedure (summarized in
Fig.~\ref{fig:sizing_flowchart_linear_final}) is: estimate correlation length
and integral range from two-point statistics; choose a high grain-size quantile
to guard against periodic self-interaction; run pilot solves to obtain
$K_{\rm app}$, $\beta_{\rm app}$, and $\lambda_{\rm app}$; enforce a length rule
on the shortest edge using microstructural, capillary, and geometric constraints;
and choose the volume to meet a target coefficient of variation, using either
the generic $(C_P A)/V$ law or the explicit capillarity prefactor, optionally
with $(\phi_\lambda,A_\lambda)$ when screening is strong. The approach is
solver-agnostic (FE or FFT), applies to image-based or synthetic media, and
turns the rule of thumb “a few particles across the box’’ into a transparent,
verifiable procedure.

Our analysis assumes small-slope capillarity, stationarity with integrable
covariance, and first-order periodic homogenization. When the covariance is
non-integrable (algebraic long-range correlations), the integral range $A$
diverges and only spectral criteria based on $k_{\min}$ and empirical variance
estimates remain predictive. In hyperuniform media with $\widehat C(\mathbf 0)=0$
one has $A=0$ and $\mathrm{Var}[P(Y)]$ decays faster than $1/V$; our formulas
then overestimate fluctuations and provide conservative lower bounds on the box
size. For large interfacial slopes, pore-scale contact-line hysteresis, or strongly heterogeneous $\lambda(\mathbf x)$ (contact-angle variations, local curvature singularities, pore-throat entry events), the screened-Helmholtz surrogate is only a coarse proxy. Accordingly, the present RVE analysis applies to the \emph{macroscopic screened capillary field} and its finite-size fluctuations, not to strongly nonlinear localized instabilities such as Haines jumps, burst events, or snap-off. In those regimes, the box sizes proposed here should be interpreted as starting scales for coarse-grained representativity, to be complemented by dedicated pore-scale or transient analysis if local instability statistics are the primary observable.

Future work will extend the validation to two-phase flow with controlled
polydispersity, link the decay length to pore–throat statistics and
experimental observables, and exploit the variance laws to optimize the
precision–cost trade-off in large-scale simulations.

\section*{Data availability}

Python script used to generate the numerical results and figures in this work, together with example input and output data, are openly available in the GitHub repository
\url{https://github.com/Christian48596/capillarity-supercell-sizing}.

\clearpage
\begin{figure*}[!t]
\centering
\footnotesize
\begin{tikzpicture}[
  node distance=4.2mm,
  box/.style={
    draw, rounded corners, align=center, fill=white,
    inner sep=3pt, text width=0.86\textwidth
  },
  arr/.style={-Stealth,very thick}
]

\node[box] (b1) {
  \textbf{1) Microstructure statistics (Section \S\ref{subsec:integral_range_variance})}\\
  Estimate correlation length $\xi$ and integral range $A$ from $C(\mathbf r)$ or $\widehat C(\mathbf k)$. Finite-size variance obeys $\mathrm{Var}[P(Y)]\!\approx (C_P A)/V$.\\
  {\footnotesize Refs: $A$ in Eq.~\eqref{eq:int_range}; variance and $1/V$ scaling Eqs.~\eqref{eq:var_phi}--\eqref{eq:cv_rule}, generic form Eq.~\eqref{eq:var_P}.}
};

\node[box, below=of b1] (b2) {
  \textbf{2) Low-wavenumber coverage (Section \S\ref{subsec:low_k_constraint})}\\
  Use the lattice $\mathcal{K}(Y)$ and require $k_{\min}^{\text{box}} = 2\pi/\min_\alpha(L_\alpha) \ll k_c$; practically $\min(L_\alpha)\gtrsim c_\xi\,\xi$.\\
  {\footnotesize Refs: Eq.~\eqref{eq:k_lattice}, criterion Eq.~\eqref{eq:lowk_criterion}, practical rule Eq.~\eqref{eq:L_vs_xi}.}
};

\node[box, below=of b2] (b3) {
  \textbf{3) Geometric safeguard for polydispersity (Section \S\ref{subsec:geom_safeguard})}\\
  Pick a high quantile $a_{\max}=Q_{1-\delta}(D)$; enforce $\min(L_\alpha)\gtrsim c_a\,a_{\max}$.\\
  {\footnotesize Refs: $a_{\max}$ in Eq.~\eqref{eq:amax_def}; safeguard Eq.~\eqref{eq:geom_guard}.}
};

\node[box, below=of b3] (b4) {
  \textbf{4) Pilot cell solves (Section \S\ref{subsec:cell_capillarity})}\\
  Solve periodic cell problems to obtain $K_{\rm app}$ and $\beta_{\rm app}$, then $\lambda_{\rm app}=\sqrt{K_{\rm iso}/\beta_{\rm app}}$, with $K_{\rm iso}=\tfrac{1}{3}\mathrm{tr}\,K_{\rm app}$.\\
  {\footnotesize Refs: Eqs.~\eqref{eq:K_app_from_cell}, \eqref{eq:beta_via_mean}--\eqref{eq:beta_average}, \eqref{eq:lambda_app_scalar}.}
};

\node[box, below=of b4] (b5) {
  \textbf{5) Length thresholds \& combination (Section \S\ref{subsec:capillarity_sizing})}\\
  Require: $\min(L_\alpha)\gtrsim c_\xi\,\xi$;\; $\min(L_\alpha)\gtrsim c_\lambda\,\lambda_{\rm app}$;\; $\min(L_\alpha)\gtrsim c_a\,a_{\max}$; anisotropic: $L_\alpha\gtrsim c_\lambda\,\lambda_{\rm app}^{(\alpha)}$ with $(\lambda_{\rm app}^{(\alpha)})^{-2}=\beta_{\rm app}/(K_{\rm app})_{\alpha\alpha}$.\\
  Combine: $L_\star=\max\{c_\xi\xi,\;c_\lambda\lambda_{\rm app},\;c_a a_{\max}\}$ and enforce aspect ratio $\min(L_\alpha)/\max(L_\alpha)\ge \rho_{\min}$.\\
  {\footnotesize Refs: Eqs.~\eqref{eq:size_micro}, \eqref{eq:size_cap}, \eqref{eq:size_cap_aniso}, \eqref{eq:Lstar}, \eqref{eq:aspect}.}
};

\node[box, below=of b5] (b6) {
  \textbf{6) Volume for target precision (Section \S\ref{subsec:capillarity_sizing})}\\
  For $n$ realizations and CV $\le\varepsilon$: $V\gtrsim \dfrac{C_P A}{n\varepsilon^2}$. For $P=\beta_{\rm app}$ (capillarity): $V\gtrsim \dfrac{c_A\xi^3}{n\varepsilon^2}\,\phi(1-\phi)\big(\tfrac{K_g-K_m}{\lambda^2}\big)^2$.\\
  {\footnotesize Refs: Eqs.~\eqref{eq:cv_rule}, \eqref{eq:V_rule_capillary_explicit}.}
};

\node[box, below=of b6] (b6b) {
  \textbf{6b) (Optional) Distribution-aware refinement (Section \S\ref{sec:capillarity_distribution_and_sizing})}\\
  Compute $\phi_\lambda$ and $A_\lambda$; optionally $\lambda_{\rm app}^{(\lambda)}$. Use $V\gtrsim \dfrac{A_\lambda}{n\varepsilon^2}\,\phi_\lambda(1-\phi_\lambda)\big(\tfrac{K_g-K_m}{\lambda^2}\big)^2$.\\
  {\footnotesize Refs: $\phi_\lambda$ Eq.~\eqref{eq:phi_lambda_def}; $A_\lambda$ Eq.~\eqref{eq:A_lambda}; rule Eq.~\eqref{eq:V_rule_distribution}; $\lambda_{\rm app}^{(\lambda)}$ via Eq.~\eqref{eq:beta_lambda_and_lambda_app}.}
};

\node[box, below=of b6b] (b7) {
  \textbf{7) Practical box choice}\\
  Cubic: $L\gtrsim \max\!\Big\{L_\star,\; \big[\dfrac{c_A\xi^3}{n\varepsilon^2}\,\phi(1-\phi)\big(\tfrac{K_g-K_m}{\lambda^2}\big)^2\big]^{1/3}\Big\}$. If using distribution-aware, replace by Eqs.~\eqref{eq:L_cubic_distribution} or \eqref{eq:L_cubic_distribution_xi}.\\
  {\footnotesize Refs: Eq.~\eqref{eq:L_cubic_capillary_explicit}.}
};

\node[box, below=of b7] (b8) {
  \textbf{8) A posteriori verification}\\
  Low-$k$: mark $k_{\min}^{\text{box}}=2\pi/\min_\alpha(L_\alpha)$ and check $k_{\min}^{\text{box}}\ll k_c$ and $k_{\min}^{\text{box}}\ll 1/\lambda_{\rm app}$;\\
  Aspect ratio satisfied; observed $\mathrm{Var}\propto 1/V$ over realizations.\\
  {\footnotesize Refs: Eqs.~\eqref{eq:lowk_criterion}, \eqref{eq:kmin_iso}, \eqref{eq:var_phi}, \eqref{eq:var_cap}, \eqref{eq:var_avg}.}
};

\draw[arr] (b1) -- (b2);
\draw[arr] (b2) -- (b3);
\draw[arr] (b3) -- (b4);
\draw[arr] (b4) -- (b5);
\draw[arr] (b5) -- (b6);
\draw[arr] (b6) -- (b6b);
\draw[arr] (b6b) -- (b7);
\draw[arr] (b7) -- (b8);

\end{tikzpicture}
\caption{Top–down workflow for choosing the periodic supercell in screened capillarity. 
\textbf{Microstructure}: estimate a correlation length and the integral range \(A\) from \(C(\mathbf r)\) or \(\widehat C(\mathbf k)\); choose a high grain--size quantile \(a_{\max}\) to avoid periodic self-interaction. 
\textbf{Physics (pilot solves)}: solve the periodic cell problems to obtain \(K_{\rm app}\) and \(\beta_{\rm app}\), then the decay length \(\lambda_{\rm app}=\sqrt{K_{\rm iso}/\beta_{\rm app}}\). 
\textbf{Length rule}: set \(L_\star=\max\{c_\xi\xi,\;c_\lambda\lambda_{\rm app},\;c_a a_{\max}\}\) and require \(\min(L_x,L_y,L_z)\gtrsim L_\star\) (use directional variants if needed). 
\textbf{Volume rule}: pick \(V\) to meet a variance target using the generic \( (C_P A)/V \) scaling or the explicit capillarity prefactor. 
\textbf{Optional distribution-aware branch}: replace \((\phi,A)\) with \((\phi_\lambda,A_\lambda)\) when screening attenuates sub--\(\lambda\) structure, and apply the corresponding variance rule. 
\textbf{A posteriori}: check low-\(k\) coverage via \(k_{\min}^{\text{box}}\) and verify the \(1/V\) decay over realizations.}
\label{fig:sizing_flowchart_linear_final}
\end{figure*}

\bibliography{main}

\end{document}